%
\let\includefigures=\iftrue
%
%
%
\newfam\black
\input harvmac.tex
\input rotate
\input epsf
\input xyv2
\noblackbox
%
%
\includefigures
\message{If you do not have epsf.tex (to include figures),}
\message{change the option at the top of the tex file.}
\def\figin{\epsfcheck\figin}\def\figins{\epsfcheck\figins}
\def\epsfcheck{\ifx\epsfbox\UnDeFiNeD
\message{(NO epsf.tex, FIGURES WILL BE IGNORED)}
\gdef\figin##1{\vskip2in}\gdef\figins##1{\hskip.5in}
\else\message{(FIGURES WILL BE INCLUDED)}%
\gdef\figin##1{##1}\gdef\figins##1{##1}\fi}
\def\DefWarn#1{}
\def\N{{\cal N}}
\def\figinsert{\goodbreak\midinsert}
\def\ifig#1#2#3{\DefWarn#1\xdef#1{fig.~\the\figno}
\writedef{#1\leftbracket fig.\noexpand~\the\figno}%
\figinsert\figin{\centerline{#3}}\medskip\centerline{\vbox{\baselineskip12pt
\advance\hsize by -1truein\noindent\footnotefont{\bf Fig.~\the\figno:} #2}}
\bigskip\endinsert\global\advance\figno by1}
\else
\def\ifig#1#2#3{\xdef#1{fig.~\the\figno}
\writedef{#1\leftbracket fig.\noexpand~\the\figno}%
\global\advance\figno by1}
\fi

\def\taueff{t}
\def\subsubsec#1{\bigskip\noindent{\it #1}}
\def\yboxit#1#2{\vbox{\hrule height #1 \hbox{\vrule width #1
\vbox{#2}\vrule width #1 }\hrule height #1 }}
\def\fillbox#1{\hbox to #1{\vbox to #1{\vfil}\hfil}}
\def\ybox{{\lower 1.3pt \yboxit{0.4pt}{\fillbox{8pt}}\hskip-0.2pt}}
\def\ttheta{{\psi}}

\def\rightarrowbox#1#2{
  \setbox1=\hbox{\kern#1{${ #2}$}\kern#1}
  \,\vbox{\offinterlineskip\hbox to\wd1{\hfil\copy1\hfil}
    \kern 3pt\hbox to\wd1{\rightarrowfill}}}

\def\cc{{\rm c.c.}}

\def\p{\partial}

\def\half{{1\over 2}}
\def\Tr{{{\rm Tr~ }}}
\def\tr{{\rm tr\ }}
\def\ad{{\rm ad~ }}
\def\Str{{\rm Str\ }}

\def\vev#1{\langle{#1}\rangle}
\def\bigvev#1{\left\langle{#1}\right\rangle}
\def\Dslash{\rlap{\hskip0.2em/}D}

\def\CC{{\cal C}}

\def\CF{{\cal F}}

\def\CN{{\cal N}}
\def\CO{{\cal O}}
\def\O{{\cal O}}
\def\CP{{\cal P}}
\def\CQ{{\cal Q}}
\def\S{{\cal S}}
\def\CS{{\cal S}}

\def\tilde{\widetilde}

\def\II{\relax{I\kern-.10em I}}

\def\bar{\overline}

\def\IZ{\relax\ifmmode\mathchoice
{\hbox{\cmss Z\kern-.4em Z}}{\hbox{\cmss Z\kern-.4em Z}}
{\lower.9pt\hbox{\cmsss Z\kern-.4em Z}}
{\lower1.2pt\hbox{\cmsss Z\kern-.4em Z}}\else{\cmss Z\kern-.4em
Z}\fi}
\def\IB{\relax{\rm I\kern-.18em B}}
\def\IC{{\relax\hbox{$\inbar\kern-.3em{\rm C}$}}}
\def\ID{\relax{\rm I\kern-.18em D}}
\def\IE{\relax{\rm I\kern-.18em E}}
\def\IF{\relax{\rm I\kern-.18em F}}
\def\IG{\relax\hbox{$\inbar\kern-.3em{\rm G}$}}
\def\IGa{\relax\hbox{${\rm I}\kern-.18em\Gamma$}}
\def\IH{\relax{\rm I\kern-.18em H}}
\def\II{\relax{\rm I\kern-.18em I}}
\def\IK{\relax{\rm I\kern-.18em K}}
\def\IN{\relax{\rm I\kern-.18em N}}
\def\IP{\relax{\rm I\kern-.18em P}}

%
\def\inbar{\,\vrule height1.5ex width.4pt depth0pt}

\def\p{\partial}

\font\cmss=cmss10 \font\cmsss=cmss10 at 7pt
\def\IR{\relax{\rm I\kern-.18em R}}

\def\id {{\bf 1}}

\def\lp10{l_P^{10}}
\def\lp11{l_P^{11}}
\def\R11{R_{11}}

%
%
\lref\VenezianoAH{ G.~Veneziano and S.~Yankielowicz, ``An
Effective Lagrangian For The Pure N=1 Supersymmetric Yang-Mills
Theory,'' Phys.\ Lett.\ B {\bf 113}, 231 (1982).
}

\lref\agmoo{ O.~Aharony, S.~S.~Gubser, J.~M.~Maldacena, H.~Ooguri
and Y.~Oz, ``Large N field theories, string theory and gravity,''
Phys.\ Rept.\  {\bf 323}, 183 (2000) arXiv:hep-th/9905111.}

\lref\NovikovEE{V.~A.~Novikov, M.~A.~Shifman, A.~I.~Vainshtein
and V.~I.~Zakharov, ``Instanton Effects In Supersymmetric
Theories,'' Nucl.\ Phys.\ B {\bf 229}, 407 (1983).
}

\lref\ArgyresJJ{ P.~C.~Argyres and M.~R.~Douglas, ``New phenomena
in SU(3) supersymmetric gauge theory,'' Nucl.\ Phys.\ B {\bf
448}, 93 (1995) [arXiv:hep-th/9505062].
}

\lref\ArgyresXN{ P.~C.~Argyres, M.~Ronen Plesser, N.~Seiberg and
E.~Witten, ``New N=2 Superconformal Field Theories in Four
Dimensions,'' Nucl.\ Phys.\ B {\bf 461}, 71 (1996)
[arXiv:hep-th/9511154].
}

\lref\SeibergRS{ N.~Seiberg and E.~Witten, ``Electric - magnetic
duality, monopole condensation, and confinement in N=2
supersymmetric Yang-Mills theory,'' Nucl.\ Phys.\ B {\bf 426}, 19
(1994) [Erratum-ibid.\ B {\bf 430}, 485 (1994)]
[arXiv:hep-th/9407087].
}

\lref\kovner{A. Kovner and M. Shifman, ``Chirally Symmetric Phase
Of Supersymmetric Gluodynamics,'' Phys. Rev. {\bf D56} (1997)
2396, hep-th/9702174.}

\lref\konishione{ K.~Konishi, ``Anomalous Supersymmetry
Transformation Of Some Composite Operators In Sqcd,'' Phys.\
Lett.\ B {\bf 135}, 439 (1984).
}

\lref\konishitwo{ K.~i.~Konishi and K.~i.~Shizuya, ``Functional
Integral Approach To Chiral Anomalies In Supersymmetric Gauge
Theories,'' Nuovo Cim.\ A {\bf 90}, 111 (1985).
}

\lref\arkmur{N. Arkani-Hamed and H. Murayama, hep-th/9707133.}

\lref\bipz{ E.~Brezin, C.~Itzykson, G.~Parisi and J.~B.~Zuber,
``Planar Diagrams,'' Commun.\ Math.\ Phys.\  {\bf 59}, 35 (1978).
}

\lref\BershadskyCX{ M.~Bershadsky, S.~Cecotti, H.~Ooguri and
C.~Vafa, ``Kodaira-Spencer theory of gravity and exact results
for quantum string amplitudes,'' Commun.\ Math.\ Phys.\  {\bf
165}, 311 (1994) [arXiv:hep-th/9309140].
}

\lref\mmreview{ P.~Ginsparg and G.~W.~Moore, ``Lectures On 2-D
Gravity And 2-D String Theory,'' arXiv:hep-th/9304011.
}

\lref\gorsky{ A.~Gorsky, ``Konishi anomaly and N = 1 effective
superpotentials from matrix models,'' arXiv:hep-th/0210281.
}

\lref\CachazoJY{ F.~Cachazo, K.~A.~Intriligator and C.~Vafa, ``A
large N duality via a geometric transition,'' Nucl.\ Phys.\ B {\bf
603}, 3 (2001) [arXiv:hep-th/0103067].
}

\lref\KutasovVE{ D.~Kutasov, ``A Comment on duality in N=1
supersymmetric nonAbelian gauge theories,'' Phys.\ Lett.\ B {\bf
351}, 230 (1995) [arXiv:hep-th/9503086].
}

\lref\FerrariJP{ F.~Ferrari, ``On exact superpotentials in
confining vacua,'' arXiv:hep-th/0210135.
}

\lref\WittenXI{ E.~Witten, ``The Verlinde Algebra And The
Cohomology Of The Grassmannian,'' arXiv:hep-th/9312104, and in
{\it Quantum Fields And Strings: A Course For Mathematicians},
ed. P. Deligne et. al. (American Mathematical Society, 1999),
vol. 2, pp. 1338-9.
}

\lref\SeibergPQ{ N.~Seiberg, ``Electric - magnetic duality in
supersymmetric nonAbelian gauge theories,'' Nucl.\ Phys.\ B {\bf
435}, 129 (1995) [arXiv:hep-th/9411149].
}

\lref\IntriligatorID{ K.~A.~Intriligator and N.~Seiberg,
``Duality, monopoles, dyons, confinement and oblique confinement
in supersymmetric SO(N(c)) gauge theories,'' Nucl.\ Phys.\ B {\bf
444}, 125 (1995) [arXiv:hep-th/9503179].
}

\lref\FujiWD{ H.~Fuji and Y.~Ookouchi, ``Comments on effective
superpotentials via matrix models,'' arXiv:hep-th/0210148.
}

\lref\KutasovNP{ D.~Kutasov and A.~Schwimmer, ``On duality in
supersymmetric Yang-Mills theory,'' Phys.\ Lett.\ B {\bf 354},
315 (1995) [arXiv:hep-th/9505004].
}

\lref\KutasovSS{ D.~Kutasov, A.~Schwimmer and N.~Seiberg,
``Chiral Rings, Singularity Theory and Electric-Magnetic
Duality,'' Nucl.\ Phys.\ B {\bf 459}, 455 (1996)
[arXiv:hep-th/9510222].
}

\lref\IntriligatorAU{ K.~A.~Intriligator and N.~Seiberg,
``Lectures on supersymmetric gauge theories and
electric-magnetic  duality,'' Nucl.\ Phys.\ Proc.\ Suppl.\  {\bf
45BC}, 1 (1996) [arXiv:hep-th/9509066].
}

\lref\DijkgraafFC{ R.~Dijkgraaf and C.~Vafa, ``Matrix models,
topological strings, and supersymmetric gauge theories,''
arXiv:hep-th/0206255.
}

\lref\DijkgraafVW{ R.~Dijkgraaf and C.~Vafa, ``On geometry and
matrix models,'' arXiv:hep-th/0207106.
}

\lref\DijkgraafDH{ R.~Dijkgraaf and C.~Vafa, ``A perturbative
window into non-perturbative physics,'' arXiv:hep-th/0208048.
}

\lref\IntriligatorJR{ K.~A.~Intriligator, R.~G.~Leigh and
N.~Seiberg, ``Exact superpotentials in four-dimensions,'' Phys.\
Rev.\ D {\bf 50}, 1092 (1994) [arXiv:hep-th/9403198].
}

\lref\SeibergBZ{ N.~Seiberg, ``Exact results on the space of
vacua of four-dimensional SUSY gauge theories,'' Phys.\ Rev.\ D
{\bf 49}, 6857 (1994) [arXiv:hep-th/9402044].
}

\lref\DijkgraafPP{ R.~Dijkgraaf, S.~Gukov, V.~A.~Kazakov and
C.~Vafa, ``Perturbative analysis of gauged matrix models,''
arXiv:hep-th/0210238.
}

\lref\GopakumarWX{ R.~Gopakumar, ``${\cal N}=1$ Theories and a
Geometric Master Field,'' arXiv:hep-th/0211100.
}

\lref\Schnitzer{S.G.~ Naculich, H.J.~ Schnitzer and N.~Wyllard,
``The $\CN=2$ $U(N)$ gauge theory prepotential and periods from a
perturbative matrix model calculation,'' arXiv:hep-th/0211123.}

\lref\SeibergVC{ N.~Seiberg, ``Naturalness versus supersymmetric
nonrenormalization theorems,'' Phys.\ Lett.\ B {\bf 318}, 469
(1993) [arXiv:hep-ph/9309335].
}

\lref\DouglasNW{ M.~R.~Douglas and S.~H.~Shenker, ``Dynamics of
SU(N) supersymmetric gauge theory,'' Nucl.\ Phys.\ B {\bf 447},
271 (1995) [arXiv:hep-th/9503163].
}

\lref\CachazoPR{ F.~Cachazo and C.~Vafa, ``N = 1 and N = 2
geometry from fluxes,'' arXiv:hep-th/0206017.
}

\lref\IntriligatorJR{ K.~A.~Intriligator, R.~G.~Leigh and
N.~Seiberg, ``Exact superpotentials in four-dimensions,'' Phys.\
Rev.\ D {\bf 50}, 1092 (1994) [arXiv:hep-th/9403198].
}

\lref\DijkgraafXD{ R.~Dijkgraaf, M.~T.~Grisaru, C.~S.~Lam,
C.~Vafa and D.~Zanon, ``Perturbative Computation of Glueball
Superpotentials,'' arXiv:hep-th/0211017.
}
\lref\superspace{ S.~J.~Gates, M.~T.~Grisaru, M.~Rocek and
W.~Siegel, ``Superspace, Or One Thousand And One Lessons In
Supersymmetry,'' Front.\ Phys.\  {\bf 58}, 1 (1983)
[arXiv:hep-th/0108200].
}

\lref\nicolai{ H.~Nicolai, ``On A New Characterization Of Scalar
Supersymmetric Theories,'' Phys.\ Lett.\ B {\bf 89}, 341 (1980).
}

\lref\migdal{ A.~A.~Migdal, ``Loop Equations And 1/N Expansion,''
Phys.\ Rept.\  {\bf 102}, 199 (1983).
}

\lref\staudacher{ M.~Staudacher, ``Combinatorial solution of the
two matrix model,'' Phys.\ Lett.\ B {\bf 305}, 332 (1993)
[arXiv:hep-th/9301038].
}

\lref\voiculescu{{\it Free Probability Theory}, ed. D. Voiculescu,
pp. 21--40, AMS, 1997.}

\lref\CeresoleZS{ A.~Ceresole, G.~Dall'Agata, R.~D'Auria and
S.~Ferrara, ``Spectrum of type IIB supergravity on AdS(5) x
T(11): Predictions on N  = 1 SCFT's,'' Phys.\ Rev.\ D {\bf 61},
066001 (2000) [arXiv:hep-th/9905226].
}

\lref\ofer{O.~Aharony, unpublished.}

\newbox\tmpbox\setbox\tmpbox\hbox{\abstractfont RUNHETC-2002-09}
\Title{\vbox{\baselineskip12pt\hbox to\wd\tmpbox{\hss
hep-th/0211170}\hbox{RUNHETC-2002-45}}}
{\vbox{\centerline{Chiral Rings and Anomalies}
\smallskip
\centerline{in Supersymmetric Gauge Theory}}}
\smallskip
\centerline{ Freddy Cachazo,$^1$ Michael R.
Douglas,$^{2,}$\footnote{$^{3}$}{ Louis Michel Professor} Nathan
Seiberg$^1$ and Edward Witten$^1$}
\smallskip
\bigskip
\centerline{$^1$School of Natural Sciences, Institute for Advanced
Study, Princeton NJ 08540 USA}
\medskip
\centerline{$^2$Department of Physics and Astronomy, Rutgers
University, Piscataway, NJ 08855-0849 USA}
\medskip
\centerline{$^3$I.H.E.S., Le Bois-Marie, Bures-sur-Yvette, 91440
France}
\bigskip
\vskip 1cm
 \noindent
 Motivated by recent work of Dijkgraaf and Vafa, we study
anomalies and the chiral ring structure in a supersymmetric
$U(N)$ gauge theory with an adjoint chiral superfield and an
arbitrary superpotential.  A certain generalization of the
Konishi anomaly leads to an equation which is identical to the
loop equation of a bosonic matrix model.  This allows us to solve
for the expectation values of the chiral operators as functions
of a finite number of ``integration constants.''  From this, we
can derive the Dijkgraaf-Vafa relation of the effective
superpotential to a matrix model.  Some of our results are
applicable to more general theories.  For example, we determine
the classical relations and quantum deformations of the chiral
ring of $\N=1$ super Yang-Mills theory with $SU(N)$ gauge group,
showing, as one consequence, that all supersymmetric vacua of
this theory have a nonzero chiral condensate.

\Date{November 2002}
%

%
%
\newsec{Introduction}

It is widely hoped that gauge theories with $\CN=1$ supersymmetry
will be relevant for real world physics. Much work has been done
on their dynamics.  In the early and mid-1990's, holomorphy of the
effective superpotential and gauge couplings were used, together
with numerous other arguments, to obtain many nonperturbative
results about supersymmetric dynamics. For a review, see e.g.
\IntriligatorAU. Many such results were later obtained by a
variety of constructions that depend on embedding the gauge
theories in string theory. These provided systematic derivations
of many results for extensive but special classes of theories.

Recently Dijkgraaf and Vafa \DijkgraafDH\ have made a striking
conjecture, according to which the exact superpotential and gauge
couplings for a wide class of $\CN=1$ gauge theories can be
obtained by doing perturbative computations in a closely related
matrix model, in which the superpotential of the gauge theory is
interpreted as an ordinary potential. Even more strikingly, these
results are obtained entirely from the {\it planar} diagrams of
this matrix model, even though no large $N$ limit is taken in the
gauge theory. This conjecture was motivated by the earlier work
\refs{\BershadskyCX\CachazoJY\CachazoPR\DijkgraafFC-\DijkgraafVW}
and a perturbative argument was given in \DijkgraafXD.

A prototypical example for their results, which we will focus on,
is the case of a $U(N)$ gauge theory with ${\cal N}=1$
supersymmetry and a chiral superfield $\Phi$ in the adjoint
representation of $U(N)$. The superpotential is taken to be
\eqn\superp{
W(\Phi)=\sum_{k=0}^n {g_k\over k+1}\Tr \Phi^{k+1}
}
for some $n$.\foot{This action is unrenormalizable if $n>3$, and
hence to quantize it requires a cutoff.  This is inessential for
us, since the cutoff dependence does not affect the chiral
quantities that we will study. Actually, by introducing
additional massive superfields, one could obtain a renormalizable
theory with an arbitrary effective superpotential \superp. For
example, a theory with two adjoint superfields $\Phi$ and $\Psi$
and superpotential $W(\Phi,\Psi)=\Tr(m\Psi^2+\Psi\Phi^2)$ would
be equivalent, after integrating out $\Psi$, to a theory with
$\Phi$ only and a $\Phi^4$ term in the superpotential.}
 If $W'(z)=g_n\prod_{i=1}^{n}(z-a_i)$, then by taking $\Phi$ to
have eigenvalues $a_i$, with multiplicities $N_i$ (which obey
$\sum_iN_i=N$), one breaks $U(N)$ to $G=\prod_iU(N_i)$; we denote
the gauge superfields of the $U(N_i)$ gauge group as $W_{\alpha
i}$. If the roots $a_i$ of $W'$ are distinct, as we will assume
throughout this paper, then the chiral superfields are all massive
and can be integrated out to get an effective Lagrangian for the
low energy gauge theory with gauge group $G$. This effective
Lagrangian is, as we explain in more detail in section 2, a
function of $S_i=-{1\over 32\pi^2}\Tr\, W_{\alpha i}W^{\alpha i}$
and $w_{\alpha_i} = {1\over 4\pi}\Tr\, W_{\alpha i}$. Explicitly,
the effective Lagrangian has the form
 \eqn\eqform{
 L_{eff}=\int d^2\theta\,  W_{eff} (S_i, w_{\alpha i}, g_k) +
 {\rm ~complex~conjugate~}+\int d^4\theta(\dots),  }
 where
 \eqn\newform{W_{eff} (S_i, w_{\alpha i}, g_k)= \omega (S_i,g_k)
 +\sum_{l,m} \taueff_{lm }(S_i,g_k)w_{\alpha l}w^{\alpha}_m.}
In \eqform, the ellipses refer to irrelevant non-chiral
interactions. Our goal, following \DijkgraafDH, is to determine
$\omega$ and $\taueff_{lm}$ (and to show that $W_{eff}$ is
quadratic in $w_{\alpha i}$, as has been assumed in writing
\eqform).

\def\hat{\widehat}
Since $U(N_i)=SU(N_i)\times U(1)_i$, it is natural for many
purposes (such as analyzing the infrared dynamics, where $SU(N_i)$
confines and $U(1)_i$ is weakly coupled) to separate the photons
of  $U(1)_i \subset U(N_i)$ from the  gluons of $SU(N_i) \subset
U(N_i)$. For this, we write
 \eqn\shatsd{S_i=\widehat S_i - {1\over 2N} w_{\alpha
 i}w^{\alpha}_i}
where $\widehat S_i = -{1\over 32 \pi ^2} \Tr \widehat W_{\alpha
i}^2$ is constructed out of the $SU(N_i)$ gauge fields $\widehat
W_{\alpha i}$. The $\widehat S_i$ are believed to behave as
elementary fields in the infrared, while the abelian fields
$w_{\alpha i}$ are infrared-free. Although this separation is
useful in understanding the dynamics, it is not very useful in
computing $W_{eff}$, mainly because $W_{eff}$ is quadratic in
$w_{\alpha i}$ if written in terms of $S_i$ and $w_{\alpha i}$,
while its dependence on $w_{\alpha i}$ is far more complicated if
it is written in terms of  $\hat S_i$ and $w_{\alpha i}$.

As we have already explained, in $L_{eff}$, $S_i$ is a bilinear
function, and $w_{\alpha i}$ a linear function, of the
superspace field strengths $W_{\alpha i}$.  So $L_{eff}$ is a
Lagrangian for a gauge theory with gauge group $G=\prod_iU(N_i)$.
It is a nonrenormalizable Lagrangian, and quantizing it requires a
cutoff, but (as in the footnote above), this is inessential for us
as it does not affect the chiral quantities.  In this paper, our
principal   goal is to determine $L_{eff}$ (modulo non-chiral
quantities), not to quantize it or to study the
$\prod_iU(N_i)$ dynamics.   In section 5, however, after
completing the derivation of $L_{eff}$, we will give some
applications of it, which generally do involve gauge dynamics. The
purpose of section 5 is to test and learn how to use $L_{eff}$,
once it has been derived.

Since it arises by integrating out only massive fields, $L_{eff}$
is holomorphic in the $S_i$ and $w_{\alpha i}$ near $S_i=0$.  A
term in $L_{eff}$ of given order in $S_i $ and $w_{\alpha i}$
arises only from {\it perturbative} contributions in $\Phi$ with a
certain number of loops. For example,  an $S^2$ term can arise
precisely in two-loop order. The number of loops for a given
contribution was determined in \DijkgraafDH, using results of
\BershadskyCX, and will be explained in section 2. A principal
result of \DijkgraafDH\ is that the perturbative contribution to
$L_{eff}$ with a given number of loops can be reproduced from a
perturbative contribution, with the same number of loops, in an
auxiliary matrix model  described there.  The auxiliary matrix
model is non-supersymmetric and has for its ordinary potential
the same function $W$ that is the superpotential of the
four-dimensional gauge theory. The original derivation of this
result made use of string theory. Our goal is to provide a direct
field theory derivation of the same result.

The basic technique is to compare the Konishi anomaly
\refs{\konishione,\konishitwo} and generalizations of it to the
equations of motion of the matrix model. In section 2, we describe
some basic facts about the problem.  We show that the general
form of the effective action follows the structure of the chiral
ring and from symmetry considerations.  In the process, we show
that the chiral ring has surprisingly tight properties. For
example, all single-trace operators in the chiral ring are at most
quadratic in $W_\alpha$, and the operator $S=-{1\over 32 \pi^2}
\Tr\,W_\alpha W^\alpha$ obeys a classical relation $S^N=0$, which
is subject to quantum deformation.  In the low energy pure
$SU(N)$ gauge theory, this relation, after quantum deformation,
implies that all supersymmetric vacua of $SU(N)$ supersymmetric
gluodynamics have a chiral condensate.
 In sections 3
and 4, we present our derivation. We construct the generalized
Konishi anomalies, which determine the quantum-corrected chiral
ring, in section 3, and we compare to the matrix model in section
4.  The computation depends upon a precise definition of the
$S_i$ given in section 2.7 in terms of gauge-invariant quantities
of the underlying $U(N)$ theory.  In section 5, we discuss in
detail some examples of applications of the results in some
simple cases.

The result that we want to establish compares the effective
superpotential of a four-dimensional gauge theory to a computation
involving  the ``same'' Feynman diagrams in a matrix model that
is a (bosonic) truncation of the zero momentum sector of the
four-dimensional gauge theory.  So one might think that the
starting point would be to compare the gauge theory to its zero
momentum sector or bosonic reduction.  If this strategy would
work in a naive form, the effective superpotential of the
four-dimensional gauge theory would be the same as the effective
superpotential of the theory obtained from it by dimensional
reduction to $n<4$ dimensions.  This is not so at all; as is clear
both from the string-based Dijkgraaf-Vafa derivation and from the
simple examples that we consider in section 2, the result that we
are exploring is specifically four-dimensional.

\subsubsec{Comparison With Gauge Dynamics}

Though gauge dynamics is not the main goal of the present paper,
it may help the reader if we summarize some of the conjectured and
expected facts about the dynamics of the low energy effective
gauge theory with gauge group $G=\prod_iU(N_i)$ that arises if we
try to quantize $L_{eff}$:

\item{(1)} It is believed to have a mass gap and confinement, and
as a result an effective description in terms of $G$-singlet
fields.

\item{(2)} For understanding the chiral vacuum states and the
value of the superpotential in them, the important singlet fields
are believed to be the chiral superfields $\widehat S_i= S_i +
{1\over 2 N_i} w_{\alpha i}w^{\alpha i})$ defined in \shatsd.

\item{(3)} Finally, it is believed that the  relevant aspects of
the dynamics of the $\widehat S_i$ can be understood by treating
the effective action  as a superpotential for elementary fields
$\widehat S_i$ and $w_{\alpha i}$, accounting for gauge dynamics
by adding to it the Veneziano-Yankielowicz superpotential
\VenezianoAH\ $\widetilde W=\sum_i N_i \hat S_i (1-\ln \hat S_i)$,
and extremizing the sum $W_{eff} = W_{eff} +\widetilde W$ with
respect to the $\widehat S_i$ to determine the vacua. In
particular, when this is done, the $\widehat S_i$ and therefore
also $S_i$ acquire vacuum expectation values, spontaneously
breaking the chiral symmetries of the low energy effective gauge
theory (and, depending on $W$, possibly spontaneously breaking
some exact chiral symmetries of the underlying theory).

These statements are on a much deeper, and more difficult, level
than the statements that we will explore in the present paper in
computing the effective action. Our results are strictly
perturbative and governed by anomalies; they are not powerful
enough to imply results such as the mass gap and confinement in
the gauge theory. The mass gap and confinement, in particular, are
needed for the formulation in (3), with the $S_i$ treated as
elementary fields, and other fields ignored, to make sense.

It might clarify things to outline the only precise derivation of
the Veneziano-Yankielowicz superpotential \VenezianoAH\ that we
know \refs{\IntriligatorJR,\IntriligatorAU}, focusing for brevity
on pure ${\cal N}=1$ gluodynamics with $SU(N)$ gauge group.  The
tree level action of the gauge theory is
 \eqn\yugoa{ L_{tree}=
 \int d^4x d^2\theta\ 2\pi i\tau_{bare} S +{c.c.} }
Logarithmic divergences of the one loop graphs force us to
replace \yugoa\ with
 \eqn\yugo{ L_{tree}= \int d^4x d^2\theta\ 3N \ln
 \left({\Lambda \over \Lambda_0}\right) S +{c.c.}}
where $\Lambda_0$ is an ultraviolet cutoff and $\Lambda$ is a
finite scale which describes the theory.  We can think of \yugo\
as a microscopic action which describes the physics at the scale
$\Lambda_0$.  The long distance physics does not change when
$\Lambda_0$ varies with fixed $\Lambda$.  The factor of $3N$
comes from the coefficient of the one-loop beta function.
Consider performing the complete path integral of the theory over
the gauge fields. Nonperturbatively, a massive vacuum is
generated, with a superpotential
 \eqn\tygos{ W_{eff}(\Lambda)= N\Lambda^3. }
Since the well-defined instanton factor is $\Lambda^{3N}$, we
prefer to write this as
 \eqn\tygoc{ W_{eff}(\Lambda)= N\left(\Lambda^{3N}\right)^{1\over
 N}. }
This superpotential controls the tension of BPS domain walls. The
$N^{th}$ root is related to chiral symmetry breaking.  The $N$
branches correspond to the $N$ vacua associated with chiral
symmetry breaking.

Now looking back at \yugo, we see that $3N\ln \Lambda$ couples
linearly to $S$ and thus behaves as a ``source'' for $S$. The
superpotential $W_{eff}(\Lambda)$ is defined directly by the
gauge theory path integral. If we want to compute an effective
superpotential for $S$, the general recipe is to introduce $S$ as
a $c$-number field linearly coupled to the source $\ln \Lambda$,
and perform a Legendre transform of $W_{eff}(\Lambda)$.  A simple
way to do that is to introduce an auxiliary field $C$ and to
consider the superpotential
 \eqn\tygoa{ W_{eff}(\Lambda, S, C)= NC^3 + S\ln
 \left({\Lambda^{3N} \over C^{3N}}\right)
 }
Integrating out $S$ using its equation of motion sets
$C^{3N}=\Lambda^{3N}$ and leads to \tygoc, showing that
$W_{eff}(\Lambda, S, C)$ of \tygoa\ leads to results identical to
those found from $W_{eff}(\Lambda)$ of \tygoc. Integrating the
auxiliary field $C$ out of \tygoa\ using its equation of motion
leads to the effective superpotential for $S$:
 \eqn\rury{W_{eff}(S)=S\left[\ln\left({\Lambda ^{3N} \over S^N}
 \right)+ N\right].
 }
This gives a clear-cut derivation and explanation of the meaning
of the Veneziano-Yankielowicz superpotential \VenezianoAH, but to
make the derivation one must already know about chiral symmetry
breaking and the nonzero $W_{eff}(\Lambda)$. One does not at present
have a derivation in which one first computes the superpotential,
proves that $S$ can be treated as an elementary field, and then uses
the superpotential to prove chiral symmetry breaking.

In this paper, we concentrate  first on generating the function
$W_{eff}(S_i)$, understood (as we have explained above in detail)
as an effective Lagrangian for the low energy gauge fields.  We do
not (until section 5)  analyze the gauge dynamics, give or assume
expectation values for the $S_i$, treat the $S_i$ or $\widehat
S_i$ as elementary fields, or extremize $W_{eff}(S_i)$ (or any
superpotential containing it as a contribution) with respect to
the $S_i$. These are not needed to {\it derive} the function
$W_{eff}(S_i)$. We stress that $S_i$ cannot be introduced by a
Legendre transform, as we did above for $S=\sum_i S_i$, because
they do not couple to independent sources.

We thus do not claim to have an {\it a priori} argument that
the gauge dynamics is governed by an effective Lagrangian for
elementary fields $S_i$. However, we will argue in section 4 that,
if it is described by such a Lagrangian, its effective
superpotential must be the sum of $W_{eff}(S_i)$ with a second
contribution $\widetilde W= \sum_i N_i S_i(1-\ln S_i) $, which
(in the sense that was just described) is believed to summarize
the effects of the low energy gauge dynamics.  (In some cases it
describes the effects of instantons of the underlying $U(N)$
theory, as we explain in section 5.)

Finally, we note that Dijkgraaf and Vafa showed that this additional
``gauge'' part of the superpotential can be extracted from the measure
of the matrix model.  Regrettably, in this paper we cast little new
light on this fascinating result.

\newsec{Preliminary Results}

In this section, we will obtain a few important preliminary
results, and give elementary arguments  for why, as claimed by
Dijkgraaf and Vafa, only planar diagrams contribute to the chiral
effective action.

\subsec{The Chiral Ring}

Chiral operators are simply operators (such as $\Tr\Phi^k$ or
$S$) that are annihilated by the supersymmetries $\bar Q_{\dot \alpha}$
of one chirality.  The product of two chiral operators is also
chiral.  Chiral operators are usually considered modulo operators
of the form $\{\bar Q_{\dot \alpha},\dots\}$.  The equivalence classes
can be multiplied, and form a ring called the chiral ring.  A
superfield whose lowest component is a chiral operator is called
a chiral superfield.

Chiral operators are independent of position $x$, up to $\bar Q^{\dot
\alpha} $-commutators.  If $\{\bar Q_{\dot\alpha},\O(x)\}=0$, then
 \eqn\mimbo{
{\p\over \p x^\mu} \O(x) = [P^\mu, \O(x)]
 = \{\bar Q^{\dot\alpha},[ Q^\alpha, \O(x)]\} .
} This implies \NovikovEE\ that the expectation value of a product
of chiral operators is  independent of each of their positions:
 \eqn\timbo{\eqalign{
 {\partial\over \partial x_1^{\alpha\dot\alpha}}\left\langle
 \O^{I_1}(x_1)\O^{I_2}(x_2) \ldots\right\rangle &
 =\left\langle\{\bar Q^{\dot \alpha},[ Q^\alpha,\O^{I_1}(x_1)]\}
 \O^{I_2}(x_2) \ldots \right\rangle \cr
 &  =- \sum_{k>1}\left\langle\left[ Q^\alpha,\O^{I_1}(x_1)\right]\ldots
 \left[ \bar Q^{\dot\alpha},\O^{I_k}(x_k)\right]\ldots\right\rangle \cr
 & =0.
 }}
Thus we can write $\vev{\prod_I\O^I(x)} = \vev{\prod_I\O^I}$
without specifying the positions $x$.

Using this invariance, we can take a correlation function of
chiral operators at distinct points, and separate the points by an
arbitrarily large distance.  Cluster decomposition then implies
that the correlation function factorizes \NovikovEE:
\eqn\factorization{
 \vev{\O^{I_1}(x_1) \O^{I_2}(x_2)\ldots
\O^{I_n}(x_n)} = \vev{\O^{I_1}} \vev{\O^{I_2}} \ldots
\vev{\O^{I_n}} .
}

There are no contact terms in the expectation value of a product
of chiral fields, because as we have just seen a correlation
function such as $\vev{\O^{I_1}(x_1) \O^{I_2}(x_2)\ldots}$ is
entirely independent of the positions $x_i$ and so in particular
does not have delta functions.  A correlation function of chiral
operators together with the upper component of a chiral superfield
 (for example $\langle \Tr\Phi^k\cdot\int d^2\theta
\Tr\,\Phi^m\rangle$) can have contact terms.

In the theory considered here, with an adjoint superfield $\Phi$,
we can form gauge-invariant chiral superfields $\Tr \Phi^k$ for
positive integer $k$.  These are all non-trivial chiral fields.
The gauge field strength $W_\alpha$ is likewise chiral, and though
it is not gauge-invariant, it can be used to form gauge-invariant
chiral superfields such as $\Tr\, \Phi^k W_\alpha$, $\Tr \Phi^k
W_\alpha \Phi^l W_\beta$, etc.  (Similar chiral fields involving
$W_\alpha$ were important in the duality of $SO(N)$ theories
\refs{\SeibergPQ,\IntriligatorID}.) Setting $k=l=0$, we get, in
particular, chiral superfields constructed from vector multiplets
only.

There is, however, a very simple fact that drastically simplifies
the classification of chiral operators.  If $\O$ is any
adjoint-valued chiral superfield, we have
 \eqn\tonno{ \left[\bar Q
 ^{\dot \alpha},D_{\alpha\dot\alpha}\O\right\} =
 \left[W_\alpha,O\right\}, }
($D_{\alpha\dot\alpha}= {D \over Dx^{\alpha\dot\alpha}}$ is the
bosonic covariant derivative) using the Jacobi identity and
definition of $W_\alpha$ plus the assumption that $\O$
(anti)commutes with $\bar Q^{\dot \alpha}$. Taking $\O=\Phi$, we
see that in operators such as $\Tr\, \Phi^k W_\alpha \Phi^m
W_\beta$, $W_\alpha$ commutes with $\Phi$  modulo $\{\bar
Q_{\dot\alpha},\dots\}$, so it suffices to consider only
operators $\Tr \Phi^n W_\alpha W_\beta$.

Moreover, taking $\O=W_\beta$ in the same identity, we learn that
 \eqn\yugoa{ \{\bar Q ^{\dot\alpha},[D_{\alpha\dot\alpha},W_\beta]
 \}=\{W_\alpha, W_\beta\}, }
so in the chiral ring we can make the substitution $W_\alpha
W_\beta \to -W_\beta W_\alpha$.  So in any string of $W$'s, say
$W_{\alpha_1} \dots W_{\alpha_s}$, we can assume antisymmetry in
$\alpha_1,\dots,\alpha_s$.  As the $\alpha_i$ only take two
values, we can assume $s\leq 2$.  So a complete list of
independent single-trace chiral operators is $\Tr\,\Phi^k$,
$\Tr\,\Phi^k W_\alpha$, and $\Tr\, \Phi^kW_\alpha W^\alpha$. This
list of operators has already been studied by
\refs{\ofer,\CeresoleZS}.

\subsec{Relations in The Chiral Ring}

We have seen that the generators of the chiral ring are of the
form $\Tr \Phi^k$, $\Tr W_\alpha \Phi^k$, $\Tr W_\alpha W^\alpha
\Phi^k$. These operators are not completely independent and  are
subject to relations.

The first kind of relation stems from the fact that $\Phi$ is an
$N\times N$ matrix.  Therefore, $\Tr \Phi^k$ with $k >N$ can be
expressed as a polynomial in $u_l=\Tr \Phi^l$ with $l\le N$
 \eqn\phikrel{ \Tr \Phi^k = \CP_k(u_1,...,u_N)}
In Appendix A, we show that the classical relations \phikrel\ are
modified by instantons for $k \ge 2N$.  This is similar to
familiar modifications due to instantons of classical relations
in the two dimensional ${\bf CP}^N$ model \WittenXI\ and in
certain four dimensional $\CN=1$ gauge theories \SeibergBZ. The
general story is that to every classical relation corresponds a
quantum relation, but the quantum relations may be different.

A second kind of relations follows from the tree level
superpotential $ W(\Phi)$.  As is familiar in Wess-Zumino models,
the equation of motion of $\Phi$
 \eqn\phieomca{ \partial_\Phi W(\Phi) = \bar D_{\dot\alpha}\bar
 D^{\dot \alpha } \bar \Phi}
shows that in the chiral ring $ \partial_\Phi W(\Phi)_{c.r.} =0$
(the subscript $c.r.$ denotes the fact that this equation is true
only in the chiral ring). In a gauge theory, we want to consider
gauge-invariant chiral operators; classically, for any $k$, $\Tr
\Phi^k \partial_\Phi W(\Phi)$ vanishes in the chiral ring. This
is a nontrivial relation among the generators. In section 3, we
will discuss in detail how this classical relation is modified by
the Konishi anomaly \refs{\konishione,\konishitwo} and its
generalizations.

We now turn to discuss interesting relations which are satisfied
by the operator $S=-{1\over 32 \pi^2} \Tr W_\alpha^2$.  We first
discuss the pure gauge $\CN=1$ theory with gauge group $SU(N)$.
We will comment about other gauge groups (such as $U(N)$) below.

The operator $S$ is subtle because it is a bosonic operator which
is constructed out of fermionic operators.  Since the gauge group
has $N^2-1$ generators, the Lorentz index $\alpha$ in $W_\alpha$
take two different values, and $S$ is bilinear in $W_\alpha$, it
follows from Fermi statistics that
 \eqn\srelationo{(S^{N^2})_{p} =0,}
so in particular $S$ is nilpotent.  We added the subscript $p$ to
denote that this relation is valid in perturbation theory. Soon
we will argue that this relation receives quantum corrections. It
is important that \srelationo\ is true for any $S$ which is
constructed out of fermionic $W_\alpha$. The latter does not have
to satisfy the equations of motion.

If we are interested in the chiral ring, we can derive a more
powerful result.  We will show that
 \eqn\srelationt{(S^N)_{p} = \{\bar Q_{\dot \alpha}, X^{\dot
 \alpha} \}}
for some $X^{\dot \alpha}, $ and therefore in the chiral ring
 \eqn\srelationth{(S^N)_{p, c.r.} = 0 }

For $SU(2)$, we can show \srelationt\ using the identity
\eqn\grelation{\Tr\,ABCD ={1\over 2}\left(\Tr AB\Tr CD + \Tr DA
\Tr BC - \Tr AC \Tr BD\right)} for any $SU(2)$ generators
$A,B,C,D$. Hence, allowing for Fermi statistics (which imply $\Tr
W_1W_1=\Tr W_2W_2=0$), we have \eqn\urelation{\Tr
W_1W_1W_2W_2=\left(\Tr W_1W_2\right)^2.} The left hand side is
non-chiral, as we have seen above, and the right hand side is a
multiple of $S^2$, so we have established \srelationth\ for
$SU(2)$.  In appendix B we extend this proof to $SU(N)$ with any
$N$, and we also show that $S^{N-1}\not= 0$ in the chiral ring.

If the relation $S^N=\{\bar Q_{\dot \alpha},X^{\dot \alpha}\}$ were an
exact quantum statement, it would follow that in any
supersymmetric vacuum, $\langle S^N\rangle=0$, and hence by
factorization and cluster decomposition, also $\langle
S\rangle=0$.  What kind of quantum corrections are possible in the
ring relation $S^N=0$?  In perturbation theory, because of
$R$-symmetry and dimensional analysis, there are no possible
quantum corrections to this relation. Nonperturbatively, the
instanton factor $\Lambda^{3N}$ has the same chiral properties as
$S^N$, so it is conceivable that instantons could modify the
chiral ring relation to $S^N={\rm constant}\cdot \Lambda^{3N}$. In
fact \NovikovEE, instantons lead to an expectation value $\langle
S^N \rangle =\Lambda^{3N}$, and therefore, they do indeed modify
the classical operator relation to
 \eqn\srelationf{S^N= \Lambda^{3N} + \{\bar Q_{\dot \alpha}
 ,X^{\dot \alpha} \} }
where in the chiral ring we can set the last term to zero.
Equation \srelationf\ is an exact operator relation in the theory.
It is true in all correlation functions with all operators. Also,
since it is an operator equation, it is satisfied in all the vacua
of the theory.  The relation $S^{N^2}=0$, which we recall is an
exact relation, not just a statement in the chiral ring, must also
receive instanton corrections so as to be compatible with
\srelationf.  To be consistent with the existence of a
supersymmetric vacuum in which $\langle S^N\rangle=\Lambda^{3N}$,
as well as with the classical limit in which $S^{N^2}=0$, the
corrected equation must be of the form $(S^N-\Lambda^{3N})
P(S^N,\Lambda^{3N})=0$, where $P$ is a homogeneous polynomial of
degree $N-1$ with a non-zero coefficient of $(S^N)^{N-1}$. We do
not know the precise form of $P$.

To illustrate the power of the chiral ring relation
$S^N=\Lambda^{3N}$, let us note that it implies that $\langle
S\rangle^N=\langle S^N\rangle = \Lambda^{3N}$ in {\it all}
supersymmetric vacua of the theory.  Indeed, the chiral ring
relation $S^N=\Lambda^{3N}$ is an exact operator relation,
independent of the particular state considered.  Thus, contrary to
some conjectures,
there does not exist a
supersymmetric vacuum of the $SU(N)$ supersymmetric gauge theory
with $\langle S\rangle=0$.

This situation is very similar to the analogous situation in the
the two dimensional supersymmetric ${\bf CP}^{N-1}$ model as well
as its generalization to Grassmannians (see section 3.2 of
\WittenXI).  In the ${\bf CP}^{N-1}$ model, a twisted chiral
superfield, often called $\sigma$, obeys a classical relation
$\sigma^N=0$; this is deformed by instantons to a quantum
relation $\sigma^N=e^{-I}$, where $I$ is the instanton action. (In
the {\it non-linear} ${\bf CP}^{N-1}$ model, $\sigma$ is bilinear
in fermions and associated with the generator of $H^2({\bf
CP}^{N-1})$; the classical relation $\sigma^N=0$ then follows
from Fermi statistics, as the fermions only have $2N-2$
components, and so is analogous to the classical relation
$S^{N^2}=0$ considered above.)

\bigskip\noindent{\it Role Of These Relations In Different
Approaches}

It is common that an operator relation which is ``always true''
in one formulation of a theory, being a kinematical relation that
holds off-shell, follows from the equation of motion in a second
formulation. In that second formulation, the given relation is not
true off-shell.  A familiar example is the Bianchi identity in two
dual descriptions of free electrodynamics. The Bianchi identity
is always true in the electric description of the theory, but it
appears as an equation of motion in the magnetic description.

This analogy leads to a simple interpretation of the
Veneziano-Yankielowicz superpotential $S(N+\ln
(\Lambda^{3N}/S^N))$.  It is valid in a description (difficult to
establish rigorously) in which $S$ is an unconstrained bosonic
field. In that description, we should not use the classical
equation $S^N= \{ \bar Q_{\dot \alpha}, X^{\dot \alpha} \}$ or its
quantum deformation $S^N= \Lambda^{3N} + \{ \bar Q_{\dot \alpha},
X^{\dot \alpha} \}$ to simplify the Lagrangian; such an equation
should arise from the equation of motion. Indeed, the equation
for the stationary points of this superpotential is $S^N=
\Lambda^{3N} $ (or zero in perturbation theory) which is the
relation in the chiral ring \srelationf.

{}From here through section 4 of the paper, we will determine an
effective action for the supersymmetric gauge theory with adjoint
superfield $\Phi$, understood as an action for low energy gauge
fields that enter via $S_i$.  Each term $\prod_i S_{i}^{k_i}$ that
we generate has a clear meaning for large enough $N_i$.  However,
for fixed $N_i$ and large enough $k_i$ could we not simplify this
Lagrangian, using for example the $U(N_i)$ relation
$S^{N_i^2+1}=0$? An attempt to do this would run into potentially
complicated instanton corrections, because actually the $S_i$
fields in the interactions $\prod_iS_i^{N_i}$  are smeared
slightly by the propagators of the massive $\Phi$ fields; in
attempting to simplify this to a local Lagrangian (where one could
use Fermi statistics to set $S^{N_i^2+1}=0$), one would run into
corrections from small instantons, so the attempt to simplify the
Lagrangian could not be separated from an analysis of gauge
dynamics. Alternatively, in section 5, we use a more powerful (but
less rigorous) approach in which it is assumed that the $\widehat
S_i$ can be treated as independent classical fields. From this
point of view, the ring relations for the $S_i$ (which are
written below in terms of $\widehat S_i$ and $w_{\alpha i}$) are
not valid off-shell and should arise from the equations of motion.

\bigskip\noindent{\it Behavior For Other Groups}

What happens for other groups?  Consider an $\CN=1$ theory  with
a simple gauge group $G$ and no chiral multiplets. The theory has
a global discrete chiral symmetry $\IZ_{2h(G)}$ with $h(G)$ the
dual Coxeter number of $G$ ($h=N$ for $SU(N)$). Numerous arguments
suggest that the $\IZ_{2h}$ discrete symmetry of the system is
spontaneously broken to $\IZ_2$, and that the theory has $h$ vacua
in which $\langle S^h \rangle =c(G) \Lambda^{3h}$ with a nonzero
constant $c(G)$.  We conjecture that at the classical level, $S$
obeys a relation $S^h=\{\bar Q_{\dot\alpha},X^{\dot\alpha}\}$, and
that instantons deform this relation to an exact operator
statement
 \eqn\srelationt{S^h= c(G) \Lambda^{3h} + \{ \bar Q_{\dot \alpha}
 , X^{\dot \alpha} \} }
for some $X^{\dot \alpha}$.  As for $SU(N)$, this statement would
imply that chiral symmetry breaking occurs in all supersymmetric
vacua.  It would be interesting to understand how the chiral ring
relation $S^h=0$ arises at the classical level, and in particular
how $h(G)$ enters.

Before ending this discussion, we will make a few comments about
the $U(N)$ gauge theory, as opposed to $SU(N)$. As we explained
in the introduction, here it makes sense to define \shatsd\
$S=\widehat S - {1\over 2 N} w_\alpha w^\alpha$ where $\widehat S$
is the $SU(N)$ part of $S$,  and the second term is the
contribution of the gauge fields of the $U(1)$ part $w_\alpha =
{1\over 4\pi} \Tr W_\alpha$. The quantum relation is $\widehat
S^N= \Lambda^{3N} + \{ \bar Q_{\dot \alpha}, X^{\dot \alpha} \}$.
Since there are only two $w_\alpha$ and they are fermionic, $S$
obeys
 \eqn\srelationfa{S^N= -\half \widehat S^{N-1}w_\alpha w^\alpha +
 \Lambda^{3N}  + \{ \bar Q_{\dot \alpha}, X^{\dot \alpha} \} =
  -\half S^{N-1}w_\alpha w^\alpha +
 \Lambda^{3N}  +  \{ \bar Q_{\dot \alpha}, X^{\dot \alpha} \} }
In vacua where $U(N)$ is broken to $\prod_i U(N_i)$, each with its
own $S_i=\widehat S_i - {1\over 2N_i} w_{i\alpha} w_i^\alpha$
these operators satisfy complicated relations. They follow from
the equation of motion of the effective superpotential
$\omega(S_i)$. What happens for $N_i=1$? In this case, one might
expect $\widehat S_i =0$. This is the correct answer classically
but it can be modified quantum mechanically to $\widehat S_i={\rm
constant}$. We will see that in more detail in section 5, where
we study the case with all $N_i=1$.  In this case instantons lead
to $\widehat S_i \not=0$. In order to describe this case in the
effective theory, one must include an independent field $\widehat
S_i$ even for a $U(1)$ factor.

\subsec{First Look at Perturbation Theory -- Unbroken Gauge Group}

Now we begin our study of the theory with the adjoint chiral
superfield $\Phi$.  For simplicity, we will begin with the case
of unbroken gauge symmetry.  So we expand around $\Phi=a$, where
$a$ is a ($c$-number) critical point of the function $W$ that
appears in the superpotential \eqn\forfun{
W(\Phi)=\sum_{k=0}^n{g_k\over k+1}\Tr\,\Phi^{k+1}. } The mass
parameter of the $\Phi$ field in expanding around this vacuum is
$m=W''(a)$.  We may as well assume that $a=0$, so $g_0=0$ and
$g_1=m$.

\ifig\oneloopdiag{(a) A one-loop diagram,
drawn in double line notation, with two gluons (wavy
lines) inserted on the same or opposite index loops. (b) A one-loop
diagram for computing $\langle \Tr \Phi^2\rangle$ in an external
field.  The ``$\times$'' represents the operator and the dotted lines
represent two external gluinos, corresponding to insertions of
$W_\alpha$.}
{\epsfbox{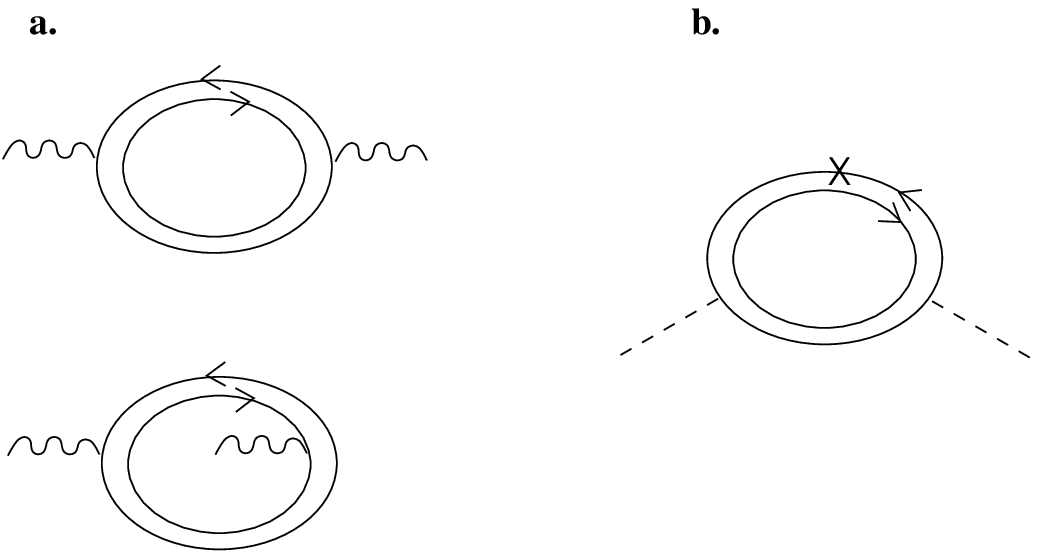}}

 Since our basic interest is in generating an effective
action for the gauge fields by integrating out $\Phi$ in
perturbation theory, we might begin with the one-loop diagram of
figure 1a. We have drawn this diagram in 't Hooft's double line
notation, exhibiting the two index lines of the $\Phi $
propagator. As is familiar, the one-loop diagram with two
external gauge fields renormalizes the gauge kinetic energy. The
result can be described (modulo non-chiral operators of higher
dimension) by the effective action
 \eqn\gorfun{ \int d^4x d^2\theta\ \ln (m/\Lambda_0)
 \left( NS + \half w_\alpha w^\alpha\right) +{c.c.}, }
where the two contributions arise from contributions where the
two gluons are inserted on the same index loop or on opposite
index loops. In the former case, one index trace gives a factor
of $N$ and the second a factor of $S=-{1\over 32\pi^2}\Tr
W_\alpha W^\alpha$; in the second case, each gives a factor of
$w_\alpha ={1\over 4\pi}\Tr W_\alpha$. \gorfun\ describes the
standard one-loop coupling constant renormalization by the
massive $\Phi$ field, with the usual logarithmic dependence on
the mass. In \gorfun, $\Lambda_0$ is an ultraviolet cutoff. The
one-loop diagram is the only contribution to the chiral couplings
that depends on $\Lambda_0$; this follows from holomorphy and
supersymmetry, or alternatively from the fact that as we will see
(following \DijkgraafDH) the multi-loop diagrams make
contributions that involve higher powers of $S$ and hence are
ultraviolet convergent.

One thing that one observes immediately is that the above result
is special to four dimensions.  If one makes a dimensional
reduction to $n<4$ dimensions before evaluating the Feynman
diagram, dimensional analysis will forbid a contribution as in
\gorfun, and a different result will ensue, depending on the
dimension. Moreover, the $m$ dependence in \forfun\ is controlled
in a familiar way by the chiral anomaly under phase rotations of
the $\Phi$ field; this is a hint that also the higher order terms
can be determined by using anomalies, as we will do in this paper.

One can phrase even the one loop results in a way in which the
cutoff does not appear, by instead computing a first derivative
of the effective superpotential with respect to the couplings. On
general grounds, these first derivatives are the expectation
values of the gauge invariant chiral operators. We will need this
result in section 4, so we will explain it in detail. A variation
of the couplings $g_k\to g_k+\delta g_k$ will produce an effective
superpotential $W_{eff}(S_i,w_{\alpha i} , g_k+\delta g_k)$ given
by
 \eqn\polyn{
 \bigvev{ \exp\left( {-\int d^4x d^2\theta\
 \sum_k {\delta g_k\over k+1} \Tr\Phi^{k+1} -c.c.}\right)
     }_{\Phi} ,
 }
where the bracket $\bigvev{ ~~~}_\Phi$ refers to the result of a
path integral over the massive fields, in the presence of a (long
wavelength) background gauge field characterized by variables
$S_i$ and $w_{\alpha i}$.  By holomorphy and supersymmetry, this
result remains valid if the coupling constants $g_k+\delta g_k$ in
\forfun\ and \polyn\ are promoted to chiral superfields
\SeibergVC. Now, take the variational derivative of \polyn\ with
respect to the {\it upper} component of $\delta g_k$. This
eliminates the $d^2\theta$ integral, and produces
 \eqn\welat{ {\partial W_{eff} \over \partial g_k}=
 \bigvev{ \Tr {\Phi^{k+1}\over k+1}}_\Phi.
 }
Thus, if we can get an expectation value as a function of the
couplings, we can integrate it to get $W_{eff}$, up to a  constant
of integration independent of the couplings.

To illustrate, let us compute the right hand side of \welat\ in
the one-loop approximation for $k=1$. The diagram that we need to
evaluate is shown in figure 1b.  There are two boson propagators
and one fermion propagator. The integral that we must evaluate is
 \eqn\toeval{ \int {d^4k\over (2\pi)^4} \left({1\over k^2+m\bar
 m}\right)^2 {\bar m\over k^2+m\bar m}={1\over 32 \pi^2 m}. }
This multiplies $NS $ or $w_\alpha w^\alpha$ depending on whether
the two external gluinos are inserted on equal or opposite index
loops, so in the one-loop approximation, we get
 \eqn\belat{
 {\partial W_{eff}\over\partial g_1}= {NS+ \half w_\alpha
 w^\alpha\over m}.
 }
(Recall that $g_1=m$.)

This illustrates the process of integrating out the massive $\Phi$
fields to get a chiral function of the background gauge fields.  To go
farther, however, we want more general arguments.

\subsec{A More Detailed Look at the Effective Action -- Unbroken
Gauge Group}

Next we will consider the multi-loop corrections.  Let us first
see what we can learn just from symmetries and dimensional
analysis. The one adjoint theory has two continuous symmetries, a
standard $U(1)_R$ symmetry and a symmetry $U(1)_\Phi$ under which
the entire superfield $\Phi$ undergoes \eqn\huffo{ \Phi\to
e^{i\alpha}\Phi. } We also introduce a linear combination of
these, $U(1)_\theta$, which is convenient in certain arguments.

If we allow the couplings to transform non-trivially,
these are also symmetries of the theory with a superpotential.
The dimensions $\Delta$
and charges $Q$ of the chiral fields and couplings are
\eqn\uonecharges{
\matrix{ & \Delta & Q_\Phi & Q_R & Q_\theta \cr
 \Phi & 1 & 1 & 2/3 &  0 \cr
 W_\alpha & 3/2 & 0 & 1 & 1 \cr
 g_l & 2-l & -(l+1) & {2\over 3} (2-l) & 2 \cr
 \Lambda^{2N}& 2N & 2N & 4N/3 & 0 \cr}
}
(Although we will not use it here,
we also included the gauge theory scale $\Lambda$ for later reference.)
Since all of these are chiral superfields,
their dimensions satisfy the relation $\Delta = 3Q_R/2$.

All of these symmetries are anomalous at one loop.  The
$U(1)_\Phi$ anomaly was first discussed by Konishi
\refs{\konishione,\konishitwo} and (as we discuss in more detail
in section 3) says that the one-loop contribution to $W_{eff}(S)$
is shifted under \huffo\ by a multiple of $S$; this can be used
to recover the result already given in \gorfun.  On the other
hand, higher loop computations are finite and the transformation
\huffo\ leaves these invariant (again, we discuss this point in
detail in section 3).  It follows that the effective action must
depend on the $g_k$'s only through the ratios
$g_k/g_1^{(k+1)/2}.$ Equivalently, we can write
 \eqn\guffo{\sum_k (k+1)g_k{\partial\over\partial g_k} W_{eff}=0,}
except at one loop.

Now let us consider dimensional analysis (by the relation
$\Delta=3Q_R/2$, $R$ symmetry leads to the same constraint). The
effective superpotential $W_{eff}$ has dimension 3. So, taking
into account the result in the last paragraph, $W_{eff}$ must
depend on the $g_k$'s and $W_\alpha$ in the form\foot{The
one-loop diagram is again an exception, because its contribution
depends on the ultraviolet cutoff $\Lambda_0$, as we see in
\gorfun, and this modifies the dimensional analysis, so again
more care is needed to study the one-loop contribution in this
way. By supersymmetry and holomorphy, $\Lambda_0$ will not enter
in the higher order contributions.}
 \eqn\inform{ W_{eff}=W_\alpha^2 F\left({g_kW_\alpha^{k-1} \over
 g_1^{(k+1)/2}} \right).
 }
Here we are being rather schematic, just to indicate the overall
power of $W_\alpha$.  So, for example, $W_\alpha^2$ may
correspond to either $S$ or $w_\alpha w^\alpha$. Instead of
\inform, we could write
 \eqn\ninform{\left(\sum_k(2-k)g_k{\partial\over\partial
 g_k}+{3\over2}W_\alpha {\partial\over\partial W_\alpha}
 \right)W_{eff}=3W_{eff}.
}

\ifig\twoloopdiag{A planar diagram with two vertices, each of $k=2$,
three index loops -- $L=3$ -- and two ordinary loops -- $L'=2$.}
{\epsffile{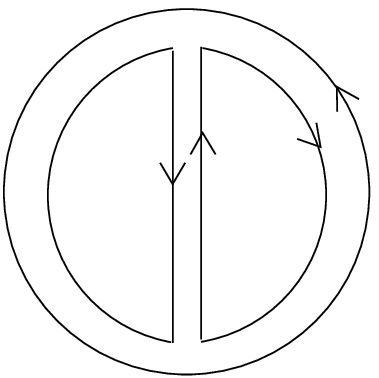}}

 If a planar Feynman diagram has vertices of degree
$k_i+1$, $i=1,\dots, s$, then the number of index loops is
\eqn\ubbu{ L=2+{1\over 2}\sum_i(k_i-1). } This is easily proved by
induction. The planar diagram sketched in figure 2 has two
vertices each with $k_i=2$; the number of index loops is 3, in
accord with \ubbu. Now each time we add a propagator to a planar
diagram, we add one index loop and increase $k_i-1$ by one at
each end of the propagator; thus, inductively, \ubbu\ remains
valid when a propagator is added or removed.

By taking a linear combination of \guffo\ and \ninform, we get
\eqn\tunno{\left(\sum_k(1-k)g_k{\partial\over\partial
g_k}+W_\alpha{\partial\over\partial
W_\alpha}\right)W_{eff}=2W_{eff}.} \ubbu\ lets us replace
$\sum_k(1-k)g_k\partial/\partial g_k$ by $4-2L$. It follows
therefore that the power of $W_\alpha$ in a contribution coming
from a planar Feynman diagram with $L$ index loops is $2L-2$. Such
a contribution might be a multiple of $S^{L-1}$, or of
$S^{L-2}w_\alpha w^\alpha$. We get $S^{L-1}$ if $L-1$ index loops
each have two insertions of external gauge bosons and one has
none, or $S^{L-2} w_\alpha w^\alpha$ if $L-2$ index loops each
have two gauge insertions and two loops have one gauge insertion
each.

The generalization of \ubbu\ for a diagram that is not planar but
has genus $g$ (that is, the Riemann surface made by filling in
every index loop with a disc has genus $g$) is
\eqn\nubbu{
L=2-2g+ {1\over 2}\sum_i(k_i-1).
}
This statement reflects 't Hooft's observation that increasing
the genus by one removes two index loops and so multiplies an
amplitude by $1/N^2$.   So  a diagram of genus $g$ has a power of
$W_\alpha$ equal to $2L+4g-2$.

At first sight, it appears that additional invariants such as
$\Tr\ W_\alpha W^\alpha W_\beta W^\beta$ will arise if there are
more then two gauge insertions on the same index loop.  But as we
have seen, such operators are non-chiral and so can be neglected
for the purposes of this paper. Also, the bilinear expression
$\Tr\ W_\alpha W_\beta$ vanishes by Fermi statistics unless the
indices $\alpha$ and $\beta$ are contracted to a Lorentz singlet.
Therefore the effective superpotential can be expressed in terms
of $S=-{1\over 32\pi^2}\Tr W_\alpha W^\alpha$ and $w_\alpha =
{1\over 4\pi}\Tr \,W_\alpha$ only.

Since each index loop in a Feynman diagram can only contribute a
factor of $S$, $w_{\alpha}$, or $N$, depending on whether the
number of gauge insertions on the loop is two, one, or zero, it
now follows, as first argued by Dijkgraaf and Vafa, that only
planar diagrams can contribute to the effective chiral
interaction for the gauge fields.  Indeed, according to \nubbu, in
the case of a non-planar diagram, at least one index loop would
have to give a trace of a product of more than two $W_\alpha$'s,
but these are trivial as elements of the chiral ring.

Furthermore, the interactions generated by planar diagrams are
highly constrained. The only way to obey \nubbu\ without
generating a trace with more than two $W$'s is to have one index
loop with no $W$ and the others with two each, or two index loops
with one $W$ and again the others with two each. The possible
interactions generated by a planar diagram with $L$ index loops
are hence $S^{L-1}$ and $S^{L-2}w_\alpha w^\alpha$.

This reproduces the results of \DijkgraafXD\ and the previous
papers cited in the introduction.

\subsec{The Case Of Broken Gauge Symmetry}

Now let us briefly consider how to extend these arguments for the
case of spontaneously broken gauge symmetry.  Here we assume that
$U(N)$ is spontaneously broken to $\prod_{i=1}^n U(N_i)$.  We
integrate out all massive fields, including the massive gauge
bosons and their superpartners, to obtain an effective action for
the massless superfields of the unbroken gauge group. The
effective action depends now on the assortment of gauge-invariant
operators $S_i=-{1\over 32\pi^2}\Tr W_{\alpha i}W^{\alpha i}$ and
$w_{\alpha i}={1\over 4\pi}\Tr W_{\alpha i}$ that appeared in the
introduction.  Their definitions will be considered more
critically in section 2.7, but for now we can be casual.

Unlike the simpler case of unbroken gauge symmetry, here we need
to integrate out massive gauge fields.  Their propagator involves
the complexified gauge coupling $\tau_{bare}$ or $\ln \Lambda$.
But the perturbative answers must be independent of its real part,
which is the theta angle, and therefore $\tau_{bare}$ cannot
appear in holomorphic quantities (except for the trivial
classical term). This means that the number of massive gauge field
propagators must be equal to the number of vertices of three
gauge fields.

The considerations of symmetries and dimensional analysis that
led to \inform\ are still valid.\foot{The one-loop diagram still
provides an exception to \inform\ because it depends
logarithmically on the cutoff $\Lambda_0$; it can be studied
specially and gives, in the usual way, a linear combination of
the $S_i$ and of $w_{\alpha i} w^\alpha_j$. This is of the
general form that the discussion below would suggest, though this
discussion will be formulated in a way that does not quite apply
at one-loop order.} However, the $W_\alpha$'s that appear in
\inform\ can now be of different kinds; they can be of the type
$W_{\alpha i}$ for any $1\leq i\leq n$. They thus can assemble
themselves into different invariants such as $S_i$ and
$w_{\alpha i}$, but for the moment we will not be specific
about this.

A key difference when the gauge symmetry is spontaneously broken
is that the dependence on the $g_k$ can be much more complicated.
The $g_k$ do not merely enter as factors at interaction vertices.
The reason is that perturbation theory is constructed by
expanding around a critical point  of $W(\Phi)$ in which the
matrix $\Phi$ has eigenvalues $a_i$ (the zeroes of $W'$) with
multiplicities $N_i$.  The $a_i$ depend on the $g_k$, as do the
masses of various components of $\Phi$; and after shifting $\Phi$
to the critical point about we intend to expand, a Feynman vertex
of order $r$ is no longer merely proportional to $g_{r-1}$ but
receives contributions from $g_s$ with $s>r-1$. Because of all
this, we need a better way to determine the relation between the
number of loops and the $S$-dependence.

Let $L'$ be the number of ordinary loops in a Feynman diagram, as
opposed to $L$, the number of index loops.  Thus, in the diagram
of figure 2, $L'=2 $ and $L=3$.  For planar diagrams, $L'=L-1$,
and in general,
 \eqn\toggo{L'=L-1+2g.}
In order to enumerate the loops we modify the tree level
Lagrangian as follows.  We replace the superpotential $W$ with
$W/\hbar$, we replace the kinetic term $\Tr \Phi^\dagger e^V
\Phi$ with ${Z \over \hbar} \Tr \Phi^\dagger e^V \Phi$, and we
replace the gauge kinetic term $\tau_{bare} W_\alpha^2$ with
${1\over \hbar} \tau_{bare} W_\alpha^2$.  Here $Z$ is a real
vector superfield, $\tau$ is a chiral superfield and $\hbar$ is
taken to be a real number. The contribution $W_{eff}^{ L'}$ to
$W_{eff}$ from diagrams with $L'$ loops is then proportional to
$\hbar^{L'-1}$, so
 \eqn\noggo{\hbar{\partial\over \partial \hbar}
 W_{eff}^{ L'}=(L'-1)W_{eff}^{ L'}.
}
Since $Z$ is a real superfield, it cannot appear in the effective
superpotential.  We have already remarked that the perturbative
superpotential is independent of $\tau_{bare}$ (except the trivial
tree dependence). Therefore, simply by calculus, since $W/\hbar$
depends only on ratios $g_k/\hbar$, we have
 \eqn\coggo{\hbar{\partial\over
 \partial \hbar} W_{eff}^{ L'}=-\sum_k g_k{\partial\over \partial
 g_k}W_{eff}^{ L'}
 }
so
\eqn\xoggo{
\sum_k g_k{\partial\over\partial
 g_k}W_{eff}^{ L'}=-(L'-1)W_{eff}^{ L'}.}
Using \toggo, the contribution $W_{eff}^{L,g}$ with $L$ index
loops and genus $g$ hence obeys \eqn\xxoggo{ \hbar{\partial\over
 \partial \hbar} W_{eff}^{ L,g}
 =-(L-2+2g)W_{eff}^{ L,g}.
}  By combining \xxoggo\ with \guffo\ and \ninform, we learn that
 \eqn\tufifo{ \sum_{\alpha, i}W_{\alpha i} {\partial\over\partial
 W_{\alpha i}}W_{eff}^{ L,g}=(2L-2+4g)W_{eff}^{ L,g},
} so a diagram with a given $L$ and $g$ makes contributions to the
superpotential that is of order $2L-2+4g$ in the $W_{\alpha i}$'s.
This is the same result as in the case of unbroken symmetry; we
have merely given a more general derivation using \xxoggo\ instead
of \nubbu. It follows from \tufifo\ that for $g>0$,  the overall
power of $W_{\alpha i}$ is greater than $2L$, so at least one
index loop contains more than two gauge insertions and gives a
trace of more than two powers of $W_{\alpha i}$ for some $i$.

Generalizing the discussion of the unbroken group, each index
loop in a Feynman diagram labeled by one of the unbroken gauge
groups $U(N_i)$ can only contribute a factor of $S_i$, $w_{\alpha
i}$, or $N_i$, depending on whether the number of gauge
insertions on the loop is two, one, or zero. It again follows
that only planar diagrams can contribute to the effective chiral
interaction for the gauge fields.  As in the unbroken case, we
learn that the possible interactions generated by a planar
diagram with $L$ index loops are $S^{L-1}$ and $S^{L-2}w_\alpha
w^\alpha$, where here each factor of $S$ or $w_\alpha$ can be any
of the $S_i$ or $w_{\alpha i}$, depending on how the index loops
in the diagram are labeled. In particular, the chiral effective
action is quadratic in the $w_{\alpha i}$.

\subsec{General Form Of The Effective Action}

Making use of the last statement to constrain the dependence on
the $w$'s,  the general form of the chiral effective action for
the low energy gauge fields is thus
 \eqn\tofigoa{
 \int d^4x\,d^2\theta\,\, W_{eff}(S_k,w_{\alpha k}) }
with
 \eqn\tofigo{
 W_{eff}(S_k, w_{\alpha k}) = \omega (S_k)+ \sum_{i,j=1}^n
\taueff_{ij}(S_k)w_{\alpha i}w^\alpha_j. }

We now make a simple observation that drastically constrains the
form of the effective action.  Since all fields in the underlying
$U(N)$ gauge theory under discussion here are in the adjoint
representation, the $U(1)$ factor of $U(N)$ is free. It is
described completely by the simple $W_{\alpha}^2$ classical
action.

In particular, there is an exact symmetry of shifting $W_\alpha$
by an anticommuting  $c$-number  (times the identity $N\times N$
matrix). In the low energy theory with gauge group $\prod_i
U(N_i)$, this symmetry simply shifts each $W_{\alpha i}$. So the
transformation of $S_i=-{1\over 32\pi^2}\Tr W_{\alpha
i}W^\alpha_i$ and $w_{\alpha i}={1\over 4\pi}\Tr W_{\alpha i}$
under $W_\alpha\to W_\alpha -4\pi \psi_\alpha$ is
 \eqn\tuffo{\eqalign{ \delta S_i&
=-\psi^\alpha w_{\alpha i}
 \cr \delta w_{\alpha_i}& =- N_i \psi_{\alpha }.
}}

It is not difficult to work out by hand the conditions imposed by
the invariance \tuffo\ on the effective action \tofigo.  However,
it is more illuminating to combine the operators $S_i$ and
$w_{\alpha i}$ in a superfield $\S_i$ which will also play a very
useful role in the sequel:
 \eqn\yuvo{ \S_i=-{1\over 2}\Tr
\left({1\over 4\pi}W_{\alpha _i}-\psi_\alpha\right)\left({1\over
4\pi}W^\alpha_i-\psi^\alpha\right) =S_i +\psi_\alpha
w^{\alpha}_{i} -
 \psi^1\psi^2N_i. }
In this description, the symmetry is simply generated by
$\partial/\partial \psi_\alpha$. Invariance under this
transformation implies that the effective action  is
 \eqn\lofog{W_{eff}=\int d^2\psi \CF_p} for some function $\CF_p$.
$\CF_p$ is not uniquely determined by \lofog, since, for example,
we could add to $\CF_p$ any function of the $SU(N_i)$ fields
$\hat S_i$ of \shatsd. However, the form of \tofigoa\ implies
that we can take $\CF_p$ to be a function only of the $\S_i$ ,
and $\CF_p$ is then uniquely determined if we also require
 \eqn\nofog{\CF_p(\S_i=0)=0.}
The subscript $p$ in the definition of $\CF_p$ denotes the fact
that  $\int d^2\psi\CF_p$ reproduces the chiral interactions
induced by integrating out the massive fields (without the
nonperturbative contribution from low energy gauge dynamics). On
doing the $\psi$ integrals, we learn that
 \eqn\plok{
 W_{eff}=\omega(S_k)+\taueff_{ij}w_{\alpha i}w^{\alpha}_j= \int
 d^2\psi
 \CF_p(\S_k)=\sum_i N_i {\partial \CF_p(S_k)\over \partial S_i} +
 \half \sum_{ij}{\partial^2 \CF_p(S_k)\over
 \partial S_i\partial S_j}w_{\alpha i}w^{\alpha}_j.}
This is the general structure claimed by Dijkgraaf and Vafa, with
$\CF_p$ still to be determined.

By expanding \plok\ in powers of $w_{\alpha i}$ (and recalling
the expansion \shatsd), we derive their gauge coupling matrix
 \eqn\gaugema{\tau_{ij} = {\partial^2 \CF_p(S_k)\over \partial S_i
 \partial S_j} -\delta_{ij} {1\over N_i} \sum_l N_l {\partial^2
 \CF_p(S_k)\over \partial S_i
 \partial S_l} .}
It is easy to see that
 \eqn\taucon{\sum_j \tau_{ij} N_j=0}
which signals the decoupling of the overall $U(1)$.  Note that it
decouples without using the equations of motion. While the
overall $U(1)$ decouples from the perturbative corrections, it of
course participates in the classical kinetic energy of the theory,
which in the present variables is $\tau\sum_iS_i$, with $\tau $
the bare coupling parameter.

In many string and brane constructions, as mentioned in
\DijkgraafDH, nonrenormalizable interactions are added to the
underlying $U(N)$ gauge theory in such a way that  the symmetry
generated by $\partial/\partial \psi$ actually becomes a second,
spontaneously broken supersymmetry.  This fact is certainly of
great interest, but will not be exploited in the present paper.
We only note that as the scale of that symmetry breaking is taken
to infinity, the Goldstone fermion multiplet which is the $U(1)$
gauge multiplet becomes free.  In this limit the broken
generators of the $\CN=2$ supersymmetry are contracted. By that
we mean that while the anticommutator of two unbroken
supersymmetry generators is a translation generator, two broken
supersymmetry generators anticommute.  Indeed, our fermionic
generators $\partial/\partial\psi$ anticommute with the analogous
anti-chiral symmetries $\partial/\partial\psi^\dagger$ that shift
$W_{\dot\alpha}$ by a constant.

Despite this relation to spontaneously broken $\CN=2$
supersymmetry in certain constructions, the symmetry \tuffo\ is
present even in theories which have no obvious $\CN=2$ analog.
For example, it would be present in any $\CN=1$ quiver gauge
theory ({\it i.e.} with $U(N)$ gauge groups and bifundamental and
adjoint matter), including chiral theories.

\subsec{$U(N)$-Invariant Description Of Effective Operators Of the
Low Energy Theory}

As preparation for the next section, we now wish to give a more
accurate description of the effective operators $S_i$ and
$w_{\alpha i}$, $i=1,\dots, n$ of the low energy theory in
terms of $U(N)$-invariant operators of the underlying $U(N)$
theory.  We assume as always that the superpotential $W$ has $n$
distinct critical points $a_i$, $i=1,\dots, n$.

Roughly speaking, we can do this by taking the $S_i$ to be linear
combinations of $\Tr\, \Phi^k W_\alpha W^\alpha$ (as we have
seen, the precise ordering of factors does not matter), for
$k=0,\dots, n-1$. Similarly, one would take the $w_{\alpha i}$
to be linear combinations of $\Tr\, \Phi^kW_\alpha$. However, to
be precise, we need to take linear combinations of these operators
to project out, for example, the $S_i$ associated with a given
critical point $a_i$.  Proceeding in this way, one would get
rather complicated formulas which might appear to be subject to
quantum corrections.

There is a more illuminating approach which also will be
extremely helpful in the next section. We let $C_i$ be a small
contour surrounding the critical point $a_i$ in the
counterclockwise direction, and not enclosing any other critical
points. And we define
\eqn\indigo{\eqalign{
S_i & = - {1\over 2\pi i}
 \oint_{C_i} dz {1\over 32\pi^2}\Tr W_\alpha W^\alpha{1\over z-\Phi} \cr
w_{\alpha_i}& ={1\over 2\pi i}\oint_{C_i} dz {1\over 4\pi} \Tr W_\alpha
{1\over z-\Phi}.\cr }}
It is easy to see, first of all, that in
the semiclassical limit, the definitions in \indigo\ give what we
want.   In the classical limit, to evaluate the integral, we set
$\Phi$ to its vacuum value. In the vacuum, $\Phi$ is a diagonal
matrix with diagonal entries $\lambda_1,\lambda_2,\dots,\lambda_N$
(which are equal to the $a_i$ with multiplicity $N_i$).  If $M$ is
any matrix and $M_{xy}$, $x,y=1,\dots, N$ are the  matrix
elements of $M$ in this basis, then
\eqn\hugb{
\Tr M{1\over z-\Phi}=\sum_x{M_{xx}\over z-\lambda_x},
}
so
\eqn\nugb{
{1 \over 2\pi i}\oint_{C_i} dz \Tr M {1\over z-\Phi}
 =\sum_x {1\over 2\pi i}\oint_{C_i} dz {M_{xx}\over z-\lambda_x}
 =\sum_{\lambda_x\in C_i} M_{xx}=\Tr\, MP_i.}
 Here $\lambda_x\in C_i$ means that $\lambda_x$ is inside the
contour $C_i$, and $P_i$ is the projector onto eigenspaces of
$\Phi$ corresponding to eigenvalues that are inside  this
contour.  In the classical limit, $P_i$ is just the projector
onto the subspace in which $\Phi=a_i$.  Hence the above
definitions amount in the classical limit to \eqn\piko{S_i
=-{1\over 32\pi^2 }\Tr\, W_\alpha W^\alpha P_i,~~w_{\alpha
i}={1\over 4\pi }\Tr W_\alpha P_i.}  This agrees with what we
want.

The formula \nugb\ is still valid in perturbation theory around
the classical limit, except that the projection matrix $P_i$ might
undergo perturbative quantum fluctuations.   One might expect
that in perturbation theory, $P_i$ in \piko\ could be expressed
as its classical value plus loop corrections that would themselves
be functions of the vector multiplets of the low energy theory. In
this case,  the objects $S_i$ and $w_{\alpha i}$ as defined in
\indigo\ would not be precisely the usual  functions of the
massless vector multiplets of the low energy theory. We will now
argue, however, that there are no such corrections. First we
complete $S_i$ and $w_{\alpha i}$ into the natural superfield
\eqn\tindigo{
\S_i=-{1\over 2\pi i}\oint dz {{1\over 2}
\Tr\,({1\over 4\pi}W_\alpha-\psi_\alpha)
    ({1\over 4\pi}W^\alpha -\psi^\alpha)}{1\over z-\Phi}.
} So \eqn\kindigo{ \S_i=S_i + \psi_\alpha w^{\alpha}_{ i} -
\psi^1\psi^2 N_i', } where \eqn\hunky{ N_i' = {1\over 2\pi
i}\oint_{C_i} dz \,\Tr{1\over z-\Phi}. }

{\it A priori}, it might seem that $N_i'$ is another chiral
superfield, but actually, in perturbation theory $N_i'$ is just
the constant $N_i$.  In fact, using \nugb, $N_i' = \Tr P_i$.
Although the projection matrix $P_i$ can fluctuate in perturbation
theory, the dimension of the space onto which it projects {\it
cannot} fluctuate, since perturbation theory only moves
eigenvalues by a bounded amount.\foot{This is trivial in
perturbation theory.  Moreover, the reduction to planar diagrams
means that perturbation theory is summable, so the rank of $P_i$
also cannot fluctuate nonperturbatively.} Thus this dimension
$N'_i$ is the constant $N_i$, the dimension of the space on which
$\Phi=a_i$. Because of this relation, we henceforth drop the
prime and refer to $N_i'$ simply as $N_i$.

Now we can show that in \piko, the $P_i$ can be simply replaced
by their classical values, so that $S_i$ and $w_{\alpha i}$ are
the usual functions of the low energy $U(N_i)$ gauge fields. The
components of the superfield $\S_i$ transform nontrivially under
the  symmetry \tuffo, but the top component of this superfield is
a $c$-number. Therefore, the bottom component $S_i$ is only a
quadratic function of the gauge fields $W_{\alpha i}$ of the low
energy theory. If fluctuations in $P_i$ caused $S_i$ to be a
non-quadratic function of the low energy $W_{\alpha i}$, then
$\partial^2 S_i/\partial \psi^2$ would be non-constant. Thus,
what is defined in \indigo\ is precisely the object that one
would want to define as $S_i$ in the low energy $U(N_i)$ gauge
theory. Likewise, the ``middle'' component $w_{\alpha i}$ of
this superfield is linear in the low energy gauge fields, so it is
precisely the superspace field strength of the $i^{th}$ $U(1)$ in
the low energy gauge theory, that is, of the center of $U(N_i)$.

In particular, since  the $S_i$ and $w_{\alpha i}$ have no
quantum corrections, they are functions only of the low energy
gauge fields and not of the $g_k$.  In section 4, we will want to
differentiate with respect to the $g_k$ at ``fixed gauge field
background''; this can be done simply by keeping $S_i$ and
$w_{\alpha i}$ fixed.

We can describe this intuitively by saying that  modulo $\{\bar
Q_{\dot\alpha},~\}$, the fluctuations in $P_i$  are pure gauge
fluctuations, roughly since there are no invariant data
associated with the choice of an $N_i$-dimensional subspace in
$U(N)$.

The meaningfulness of the  couplings $t_{ij}$ in \eqform\ depends
on having correctly normalized the field strengths $w_{\alpha i}$;
in the absence of a preferred normalization, by making arbitrary
field redefinitions one could always set
$\taueff_{ij}=\delta_{ij}$. Since the low energy gauge theory
contains particles charged under $U(1)^n$ (the $W$ bosons of
broken gauge symmetry), the preferred normalization is the one in
which these particles have integer charge in each $U(1)$ factor.
Again, this is clearly true of \indigo\ in the semiclassical
limit, and in the form \piko\ is clearly true in general.

These properties are not preserved under general superfield
redefinitions $f_j(\CS_i)$.  By analogy with conventional $\CN=2$
supersymmetry, in which one speaks of the preferred $\CN=2$
superfields for which $U(1)^n$ charge quantization is manifest as
defining ``special coordinates,'' we can also use the term special
coordinates for the variables $S_i$ we have discussed here.

\newsec{The Generalized Konishi Anomaly}

The Konishi anomaly is an anomaly for the (superfield)
current\foot{ Here $\ad V$ signifies the adjoint representation,
{\it i.e.} $(\ad V~ \Phi)^i{}_j = V^i{}_k \Phi^k{}_j - \Phi^i{}_k
V^k{}_j$.}
\eqn\Lzerocur{ J = \Tr \bar\Phi e^{\ad V} \Phi , }
which generates the infinitesimal rescaling of the chiral field,
\eqn\rescaling{ \delta\Phi =\epsilon \Phi . }
It can be computed
by any of the standard techniques: point splitting, Pauli-Villars
regularization (since our model is non-chiral), anomalous
variation of the functional measure (at one loop) or simply by
computing Feynman diagrams \refs{\konishitwo,\superspace}. The
result is a superfield generalization of the familiar
 $U(1) \times SU(N)^2$ mixed chiral
anomaly for the fermionic component of $\Phi$; in the theory with
zero superpotential,\foot{In a general supersymmetric theory, we
have to express such an  identity in terms of commutators with
$\bar Q_{\dot\alpha}$ In the present theory, we use the existence
of a superspace formalism and write the identity in terms of
$\bar D_{\dot \alpha}$.  Note that when a superspace formalism
exists, $\bar Q_{\dot\alpha}$ and $\bar D_{\dot\alpha}$ are
conjugate (by $\exp(\bar\theta\theta\partial/\partial x$), so the
chiral ring defined in terms of operators annihilated by $\bar
Q_{\dot\alpha}$ is isomorphic to that defined using $\bar
D_{\dot\alpha}$. }
 \eqn\oneloopan{
 \bar D^2 J = {1\over 32\pi^2} {\rm tr}_{adj}\ (\ad W_\alpha)(\ad
 W^\alpha) }
where the trace is taken in the adjoint representation.

Evaluating this trace and adding the classical variation present
in the theory with superpotential, we obtain
\eqn\konishian{
\bar D^2 J = \Tr \Phi {\p W(\Phi)\over\p \Phi}
 + {N\over 16\pi^2} \Tr W_\alpha W^\alpha
 - {1\over 16\pi^2} \Tr W_\alpha \Tr W^\alpha  .
}
One way to see why this combination of traces appears
here is to check that the diagonal $U(1)$ subgroup of the gauge group
decouples.

We now take the expectation value of this equation. Since the
divergence $\bar D^2 J$ is a $\bar Q^{\dot\alpha}$-commutator, it
must have zero expectation value in a supersymmetric vacuum.
Furthermore, we can use  \factorization\ and $\vev{\Tr
W_\alpha}=0$ to see that the last term is zero.  Thus, we infer
that
 $$ \left\langle\Tr \Phi {\p W(\Phi)\over\p \Phi}\right\rangle=
 - 2 {N\over 32\pi^2} \left\langle \Tr W_\alpha W^\alpha
 \right\rangle .
 $$
Using \welat, this can also be formulated in terms of a constraint
on $W_{eff}$.  The left hand side reproduces \guffo, so we find
that
 \eqn\Lzerogauge{ \sum_{k\ge 0}
 (k+1)g_k{\partial\over\partial g_k} W_{eff}= 2 N S }
where (in the notations of section 2)
 $$
 S = \sum_i S_i =  -{1\over 32\pi^2} \vev{ \Tr W_\alpha W^\alpha } .
 $$
This has the solution \gorfun, and the general solution is
\gorfun\ plus a solution of the homogeneous equation \guffo.

In general, this anomaly receives higher loop contributions,
which are renormalization scheme dependent and somewhat
complicated.  However, one can see without detailed computation
that these contributions can all be written as non-chiral local
functionals.

To see this, we consider possible chiral operator corrections to
\konishian, and enforce symmetry under the $U(1)_\Phi\times
U(1)_\theta$ of \uonecharges. An anomaly must have charges $(0,2)$
under $U(1)_\Phi\times U(1)_\theta$.
 Furthermore the correction must vanish for $g_k=0$, so inverse
powers of the couplings cannot appear. Referring to \uonecharges,
we see that the only expressions with the right charges are
$W_\alpha^2$ and $g_k\Phi^{k+1}$, with no additional dependence
on couplings. These are the terms already evaluated in
\konishian, and thus no further corrections are
possible.\foot{The situation in the full gauge theory is less
clear.  Including nonperturbative effects, one might generate
corrections such as $g_{2Nl+k-1}\Lambda^{2Nl}\Tr\Phi^k$. Such
terms would not affect the perturbative analysis of $W_{eff}$ in
section 4, but would affect the study of gauge dynamics in
section 5.  In particular, this simple form of the Ward identities
depends on making the proper definition of the operators $\Tr
\Phi^k$ for $k\ge N$, which has interesting aspects discussed in
section 5 and Appendix A. We suspect, but will not prove here,
that such nonperturbative corrections can be excluded (after
properly defining the operators) by considering the algebra of
chiral transformations $\delta\Phi = \sum_n\epsilon_n\Phi^{n+1}$
(as is well known in matrix theory, this algebra is a partial
Virasoro algebra), showing that it can undergo no quantum
corrections, and using this algebra to constrain the anomalies,
along the lines of the Wess-Zumino consistency condition for
anomalies.} This implies that, for purposes of computing the
chiral ring and effective superpotential, the result \konishian\
is exact.\foot{We can also reach this conclusion by quite a
different argument. Pauli-Villars regularization of the $\Phi$
kinetic energy improves the convergence of all diagrams
containing $\Phi$ fields with more than one loop, without
modifying chiral quantities.  So anomalies obtained by
integrating out $\Phi$ must arise only at one-loop order.}

Thus we have an exact constraint on the functional $W_{eff}$.
Readers familiar with matrix models will notice the close
similarity between \Lzerogauge\ and the $L_0$ Virasoro constraint
of the one bosonic matrix model (this was also pointed out in
\gorsky).  Indeed, the similarity is not a coincidence as the
matrix model $L_0$ constraint has a very parallel origin: it is a
Ward identity for the matrix variation $\delta M = \epsilon M$.
In the matrix model, one derives further useful constraints (which
we recall in section 4.2) from the variation $\delta M=\epsilon
M^{n+1}$.

This similarity suggests that we try to get more constraints on
the functional $W_{eff}(g_n,S_i)$ by considering possible
anomalies for the currents
 \eqn\Lncur{ J_n = \Tr \bar\Phi e^{\ad V} \Phi^{n+1} , }
which generate the variations
 \eqn\rescalings{ \delta\Phi =\epsilon_n \Phi^{n+1} . }
The hope would be that these would provide an infinite system of
equations analogous to \Lzerogauge, which could determine the
infinite series of expectation values $\vev{\Tr \Phi^n}$.

It is easy to find the generalization of the classical term
in \konishian, by applying the variations \rescalings\ to the
action.  This produces
\eqn\genclassan{
\bar D^2 J_n = \Tr \Phi^{n+1} {\p W(\Phi)\over \p\Phi}
 + {\rm anomaly} + \bar D(\ldots).
}

Following the same procedure which led to \Lzerogauge, we obtain
 \eqn\Lnclass{ \sum_{k\ge 0} (k+n+1)g_k{\partial\over\partial
 g_{k+n}} W_{eff}= {\rm anomaly\ terms} .
 }
These are independent constraints on $W_{eff}$, motivating us to
continue and compute the anomaly terms.  However, before doing
this, we should recognize that we did not yet consider the most
general variation in the chiral ring.  This would have been
 \eqn\generalvar{\delta\Phi =f(\Phi,W_\alpha)
 }
for a general holomorphic function $f(\Phi,W_\alpha)$.  It is no harder
to compute the anomaly for the corresponding current
\eqn\generalcur{
J_f = \Tr \bar\Phi e^{\ad V} f(\Phi,W_\alpha) ,
}
so let us do that.

\subsec{Computation of the generalized anomaly}

We want to compute
\eqn\genanomaly{
\bar D^2 \vev{J_f} =
\bar D^2 \vev{\Tr \bar \Phi e^{\ad V} f(\Phi,W_\alpha)} .
}
Let us first do this at zeroth order in the couplings $g_k$
(except that we assume a mass term); we will
then argue that holomorphy precludes corrections depending on these
couplings.

At zero coupling, the one loop contributions to \genanomaly\
come from graphs
involving $\bar \Phi$ and a single $\Phi$ in $f(\Phi,W_\alpha)$.
In any one of these graphs,
the other appearances of $\Phi$ and $W_\alpha$ in $f$ are simply
spectators.  In other words, given
$$
 A_{ij,kl} \equiv \bar D^2 \vev{ \bar \Phi_{ij} e^{\ad V} \Phi_{kl} } ,
$$
the generalized anomaly at one loop is
\eqn\genn{
\bar D^2 \vev{\Tr \bar \Phi e^{\ad V} f(\Phi,W_\alpha)}
 = \sum_{ijkl} A_{ij,kl} {\p f(\Phi,W_\alpha)_{ji} \over \p \Phi_{kl}} .
 }
The index structure on the right hand side should be clear on
considering the following example,
 $$ {\p \over\p \Phi_{kl}}\left(\Phi^{m+1}\right)_{ji}
 = \sum_{s=0}^{m} (\Phi^s)_{jk} (\Phi^{m-s})_{li} . $$
In general, the expression \genn\ might need to be covariantized
using $e^V$ and $e^{-V}$. However, we are only interested in the
chiral part of the anomaly, for which this is not necessary.

In fact, the computation of $A_{ij,kl}$ is the same as the
computation of \oneloopan, with the only difference being that
we do not take the trace.  Thus
$$\eqalign{
A_{ij,kl}
&= {1\over 32\pi^2}
\left[ (W_\alpha W^\alpha)_{il}\delta_{jk} +
(W_\alpha W^\alpha)_{jk}\delta_{il} -
 2 (W_\alpha)_{il}(W^\alpha)_{jk} \right]  \cr
 &\equiv {1\over 32\pi^2} \left[W_\alpha,
 \left[W^\alpha, e_{lk}\right]\right]_{ij}
}$$
where $e_{lk}$ is the basis matrix with the single non-zero
entry $(e_{lk})_{ij} = \delta_{il} \delta_{jk}$.

Substituting this in \genn\ and
adding the classical variation, we obtain the final result
\eqn\genkonishian{
\bar D^2 J_f = \Tr f(\Phi,W_\alpha) {\p W(\Phi)\over \p \Phi} +
 {1\over 32\pi^2}
 \sum_{i,j} \left[W_\alpha, \left[W^\alpha, {\p f\over\p \Phi_{ij}}\right]
 \right]_{ji} .
}

Finally, we argue that this result cannot receive perturbative (or
nonperturbative) corrections in the coupling.  The argument
starts off in the same way as for the standard Konishi anomaly,
using $U(1)_\Phi \times U(1)_\theta$ symmetry to conclude that
the only allowed terms are those involving the same powers of the
fields and the couplings as already appear in \genkonishian. The
same sort of analysis that we used in section 2 to show that
particular interactions only arise from diagrams with a
particular number of loops (for example, $S^2$ only arises from
two-loop diagrams) then shows that the terms proportional or not
proportional to $W_\alpha^2$ in \genkonishian\ can only arise
from one-loop or tree level diagrams.  But the tree level
contribution is classical, and we have already evaluated the
one-loop contribution to arrive at \genkonishian.

Taking expectation values, we obtain the Ward identities
\eqn\betterfun{
\bigvev{ \Tr f(\Phi,W_\alpha) {\p W\over\p\Phi} } =
 - {1\over 32\pi^2}
\bigvev{ \sum_{i,j} \left( \left[W_\alpha,\left[W^\alpha,
         {\p f(\Phi,W_\alpha) \over \p \Phi_{ij}}
           \right]\right]\right)_{ij} }.
}

\subsubsec{Functional measure description}

In the next section, we will use these Ward identities to solve
for $W_{eff}$.  They have a close formal similarity with those of
the one matrix model, and a good way to see the reason for this
is to rederive them using the technique of anomalous variation of
the functional measure.  This technique can be criticized on the
grounds that it is not obvious how to extend it beyond one loop,
and this is why we instead gave the more conventional
perturbative argument for \genkonishian.  Now that we have the
result in hand, we make a brief interlude to explain it from this
point of view.

A simple
way to derive classical Ward identities such as \genclassan\ is
to perform the variation under the functional integral: we write
\eqn\naivefun{\eqalign{
&\int [D\Phi] \Tr \left( f(\Phi) {\p\over\p\Phi} \right)
 \ \exp\left(-\int d^2\theta W(\Phi) - \cc\right) \cr
&= - \int d^2\theta\ \bigvev{ \Tr f(\Phi) {\p W\over\p\Phi} }. }}
In the classical theory, this would vanish by the equations of
motion. However, in the quantum theory, the functional measure
might not be invariant under such a variation. Formally, we might
try to evaluate this term by integrating by parts under the
functional integral \naivefun. This would produce the anomalous
Ward identity \eqn\anomwi{ \bigvev{ \Tr f(\Phi) {\p W\over\p\Phi}
} = \bigvev{ \sum_{i,j} {\p f(\Phi)_{ij} \over \p \Phi_{ij}} }. }
Such a term might be present in bosonic field theory, and is
clearly present in zero dimensional field theory ({\it i.e.} the
bosonic one matrix integral), as integration by parts is obviously
valid.  In fact it is simply the Jacobian for an infinitesimal
transformation $\delta \Phi = f$, \eqn\defjac{ \log J = \tr \log
\left(1 + {\delta f(\Phi) \over \delta \Phi} \right) , } which
expresses the change in functional measure from $\Phi$ to $\Phi +
\delta \Phi$.  Thus the Ward identities \anomwi, which can serve
as a starting point for solving the matrix model (and which we
will discuss further in section 4.2), in fact express this
``anomalous'' or ``quantum'' variation of the measure.

One can make the same computation in supersymmetric theory, by
computing the product of Jacobians for the individual component fields.
As is well-known (this is used, for example, in the discussion of the
Nicolai map \nicolai), the result is $J=1$, by cancellations between the
Jacobians of the bosonic and fermionic components.
Consistent with this, the anomalous term in \anomwi\ is
not present in \betterfun.

However, to properly compute the variation of the functional measure
in field theory, one must regulate the trace appearing in \defjac.
This is how the Konishi anomaly was computed in \konishitwo.
It is straightforward to generalize their computation by taking
\eqn\sfieldjac{
\eqalign{
\log J &= \epsilon\ \Str e^{-t L}
 {\delta f(\Phi,W_\alpha) \over \delta \Phi} \cr
&= \epsilon \left({\rm Tr}_\phi\ e^{-t(D_\mu)^2} {\delta
f(\Phi,W_\alpha) \over \delta \Phi}
       - {\rm Tr}_\psi\ e^{-t \Dslash^2} {\delta f(\Phi,W_\alpha) \over \delta \Phi}
    + {\rm Tr}_F\ e^{-t (D_\mu)^2} {\delta f(\Phi,W_\alpha) \over \delta \Phi}\right) ,
 }}
where $L=\bar D^2 e^{-2V} D^2 e^{2V}$ is an appropriate
superspace wave operator, to obtain precisely the result
\betterfun.

\subsubsec{More general quiver theories}

Without going into details, let us mention that all of the
arguments we just gave generalize straightforwardly to a general
gauge theory with a product of classical gauge groups,
bifundamental and adjoint matter.  An important point here is
that the identities \tonno\ and \yugoa, which freed us from the
need to specify the operator ordering of $W_\alpha$ insertions,
now free us from the need to specify which gauge group each
$W_\alpha$ lives in (in any given ordering of $\Phi^a$ and
$W_\alpha$'s, this is fixed by gauge invariance). One still needs
to keep track of the ordering of the various chiral multiplets,
call them $\Phi^a$.

One can then use a general holomorphic variation $$ \delta \Phi^a
= f^a(\Phi,W_\alpha) $$ in the arguments following \generalvar, to
derive  Ward identities very similar to \betterfun.  These are
analogs of standard Ward identities known for multi-matrix models
\staudacher\ which generalize the ``factorized loop equations''
of large $N$ Yang-Mills theory \migdal.

\newsec{Solution of The Adjoint Theory}

We now have the ingredients we need to find $W_{eff}(S,g)$.
Readers familiar with matrix models will recognize that much of
the formalism used in the following discussion was directly
borrowed from that theory.  However, we do not assume familiarity
with matrix models, nor do we rely on any assumed relation
between the problems in making the following arguments.

We also stress that the arguments in this section generally do not
assume the reduction to planar diagrams which we argued for in
section 2, but will lead to an independent derivation of this
claim. Indeed, the arguments of section 3 and this section do not
use any diagrammatic expansion and are completely
non-perturbative.

We are still considering the one adjoint theory with
superpotential \superp.  We assemble all of its chiral operators
into a generating function,
 \eqn\defCR{ \CR(z,\ttheta) = -\half
\Tr { \left({1\over 4\pi}W_\alpha-\ttheta_\alpha\right)^2}{1 \over
z-\Phi} . } We simply regard this as a function whose expansion in
powers of $1/z$ and $\psi$ generates the chiral ring. Its
components in the $\psi$ expansion are \eqn\CRcomp{ \CR(z,\ttheta)
=
 R(z) + \ttheta_\alpha w^\alpha(z) - \psi^1\psi^2 T(z)
} with \eqn\defgen{\eqalign{ T(z) &= \sum_{k\ge 0} z^{-1-k} \Tr
\Phi^k = \Tr {1\over z-\Phi} ; \cr w_\alpha(z) &= {1\over 4\pi}\Tr
W_\alpha{1\over z-\Phi} ; \cr R(z) &= -{1\over 32\pi^2}\Tr
{W_\alpha W^\alpha}{1\over z-\Phi} . }} We will also use the same
symbols to denote the vacuum expectation values of these
operators; the meaning should be clear from the context. Finally,
we write \eqn\defCRmat{ \CR(z,\ttheta)_{ij} =
 -\half
 \left({ \left({1\over 4\pi}W_\alpha-\ttheta_\alpha\right)^2}
 {1 \over z-\Phi}\right)_{ij}
}
to denote the corresponding matrix expressions (before taking the trace).

Note that the expansion in powers of $\psi$ runs ``backwards''
compared to a typical superspace formalism: the lowest component
is $R(z)$, the gaugino condensate, while operators constructed
only from powers of $\Phi$ are upper components.  Because of this,
it will turn out that $R(z)$ is the basic quantity from which the
behavior of the others can be inferred.

\subsubsec{Basic loop equations for gauge theory}

Let us start by deriving Ward identities for the lowest component
$R(z)$ in \CRcomp, i.e. with $\psi=0$.  By then taking non-zero $\psi$,
we will obtain the Ward identities following from
the most general possible variation which is a function of chiral operators.

So, we start with
\eqn\varRgen{
\delta \Phi_{ij} = f_{ij}(\Phi,W_{\alpha }) =
 -{1\over 32\pi^2} \left({W_\alpha W^\alpha \over z-\Phi}\right)_{ij} .
}
Substituting into \betterfun, one finds the Ward identity
\eqn\loopR{\eqalign{
 \bigvev{
  -{1\over 32\pi^2} \sum_{i,j} \left[W_\alpha,\left[W^\alpha,
  {\p\over\p \Phi_{ij}}
     {W_\alpha W^\alpha \over z-\Phi}\right]\right]_{ij}}
 = \bigvev{ \Tr \left({W'(\Phi) W_\alpha W^\alpha \over z-\Phi} \right)}  .
}}

We next note the identity \eqn\masteridentity{ \sum_{i,j}
\left[\chi_1,\left[\chi_2, {\p\over\p \Phi_{ij}}
 {\chi_1 \chi_2\over z-\Phi}\right]\right]_{ij} =
 \left(\Tr {\chi_1 \chi_2\over z-\Phi} \right)^2
}
which is valid if $\chi_1^2 = \chi_2^2 = 0$ and
the chiral operators $\Phi$ and
$\chi_\alpha$ commute, as can be verified by elementary algebra.

Applying this identity with
$\chi_\alpha=W_\alpha$, we find
$$
\bigvev{R(z) ~ R(z)} =
 \bigvev{\Tr \left(W'(\Phi)R(z)\right) }.
$$

We now come to the key point.  Expectation values of products
of gauge invariant chiral operators factorize, as expressed in
\factorization.  This allows us to rewrite \loopR\ as
$$
\vev{R(z)}^2 =  \bigvev{\Tr \left(W'(\Phi) R(z)\right) }.
$$
Thus, both sides have been expressed purely in terms of the
vacuum expectation values $\vev{\Tr W_\alpha W^\alpha \Phi^k}$,
allowing us to write a closed equation for $R(z)$.

This point can be made more explicit by rewriting the right hand
side in the following way. We start by noting the identity
 \eqn\mmminus{\eqalign{ {\Tr {W'(\Phi)W_\alpha W^\alpha
 \over z-\Phi}} &= {\Tr {(W'(\Phi)-W'(z)+W'(z))W_\alpha
 W^\alpha\over z-\Phi}} \cr &= W'(z) {\Tr
{W_\alpha W^\alpha\over z-\Phi}} + {1 \over 4}f(z) }} with
 $$
 f(z) = 4\ {\Tr {(W'(\Phi )-W'(z ))W_\alpha W^\alpha\over
 z-\Phi}}.
 $$
(The factor of $4$ is introduced for later convenience.) The
function $f(z)$ is analytic and, for polynomial $W$, is a
polynomial in $z$ of degree $n-1=\deg W'-1$, simply because
$W'(\Phi)-W'(z)$ vanishes at $z-\Phi=0$. In terms of this function,
we can write \eqn\mmRloop{ R(z)^2 = W'(z) R(z) + {1\over 4}f(z) }
This statement holds as a statement in the chiral ring, and
(therefore) also holds if the operators $R(z)$ and $f(z)$ are
replaced by their values in a supersymmetric vacuum.

Another way to understand this result is to note that
 $$
 \Tr {{W'(\Phi)W_\alpha W^\alpha \over z-\Phi}} \sim O(1/z)
 $$
for large $z$, while $f(z)$ as defined above is polynomial.  Thus
we can write
 $$
 -{1\over 32 \pi^2} {\Tr {W'(\Phi)W_\alpha W^\alpha\over
 z-\Phi}} = [W'(z) R(z)]_-
 $$
where the notation $[F(z)]_-$ means to drop the non-negative
powers in a Laurent expansion in $z$, {\it i.e.}
 $$ [z^k]_- =
 \cases{ z^k & for $k<0$ \cr  0 & for $k\ge 0$}.
 $$
Using this notation, we can write the  Ward identity as
 \eqn\mmRlooptwo{ R(z)^2 =  [W'(z) R(z)]_- . }
In other words, the role of $f(z)$ in \mmRloop\ is just to cancel
the polynomial part of the right hand side.

Readers  familiar with matrix models will recognize \mmRloop\ and
\mmRlooptwo\ as the standard loop equations for the one bosonic
matrix model.  If the resolvent of the matrix field $M$ is defined
by
  \eqn\resdef{R_M(z)={g_m\over \widehat N}\bigvev{ {\rm Tr}
 {1\over z-M} },}
then it obeys precisely the above Ward identities, as we recall
more fully in section 4.2. Thus, we will soon be in a position to
prove a precise relation between gauge theory expectation values
$\vev{\ldots}_{g.t.}$ and matrix model expectation values
$\vev{\ldots}_{m.m.}$,
 $$ -{1\over 32\pi^2} \bigvev{\Tr {W_\alpha W^\alpha\over
 z-\Phi}}_{g.t.}
 = {g_m\over \widehat N} \bigvev{\Tr {1\over z-M}}_{m.m.} .
$$

Let us continue, however, without assuming any {\it a priori}
relationship, and complete the computation of $W_{eff}$ purely in
terms of gauge theory.

\subsubsec{General loop equations for gauge theory}

The general system of Ward identities is no harder to derive.
We apply the general variation \generalvar\ with
\eqn\vargen{
\delta \Phi_{ij} = f_{ij}(\Phi,W_{\alpha }) = \CR(z,\ttheta)_{ij} .
}
Substituting into \betterfun, one finds the Ward identity
\eqn\loophigher{\eqalign{
 \bigvev{
  {1\over 4(4\pi)^2} \sum_{i,j} \left[W_\alpha,
  \left[W^\alpha, {\p\over\p \Phi_{ij}}
     {({1\over 4\pi}W_\alpha - \ttheta_\alpha)^2 \over z-\Phi}
     \right]\right]_{ij}}
 = \left\langle \Tr \left(W'(\Phi) \CR(z,\ttheta)\right)\right\rangle  .
}}

Now, one can substitute $\ad ({1\over 4\pi }W_\alpha - \psi_\alpha)$ for
$\ad {1\over 4\pi}W_\alpha$ in this expression, as $\ad \id = 0$.
One can then apply \masteridentity\ with
$\chi_\alpha={1\over 4\pi }W_\alpha - \psi_\alpha$, and gauge
theory factorization \factorization, exactly as before, to find
 $$
 \CR(z,\ttheta)^2 =  {\Tr \left(W'(\Phi)\CR(z,\ttheta)\right) }.
 $$
Note that the variable $\ttheta$ does not appear explicitly in
this result.  This expresses the decoupling of the diagonal
$U(1)$, and thus was clear {\it a priori}. Again, this statement
holds as a statement about the chiral ring, or as a statement
about expectation values.

 Finally, we can apply
the same identity \mmminus, now using a degree $n-1$ polynomial
$f(z,\psi)$, to obtain \eqn\mmloop{ \CR(z,\psi)^2 =  W'(z)
\CR(z,\psi) + {1\over 4}f(z,\psi) . } This equation contains
\mmRloop\ and higher components which will allow us to determine
the expectation values of the other operators in the chiral ring.
Expanding \mmloop\ in powers of $\psi$, and writing \eqn\expf{
f(z,\psi) = f(z) + \psi_\alpha \rho^\alpha(z) - \psi_1\psi_2 c(z),
} we obtain the component equations \eqn\loopcomponents{\eqalign{
R^2(z) &= W'(z) R(z) + {1\over 4}f(z) , \cr 2 R(z) w_\alpha(z) &=
W'(z) w_\alpha(z) + {1\over 4}\rho_\alpha(z) , \cr 2 R(z) T(z) +
w_\alpha(z) w^\alpha(z) &= W'(z) T(z) + {1\over 4}c(z) . }}

This system of equations is the complete set of independent Ward
identities which can be derived using the generalized Konishi
anomaly.\foot{There are further equations which can be derived by
considering correlation functions of \generalcur\ with other
chiral operators, but these are equivalent to derivatives of
\mmloop\ with respect to the couplings.} It can be used in
various ways.  For example, by expanding in powers of $1/z$, one
can derive a system of recursion relations determining all vevs
in terms of those with $k\le n$.  One can also expand $\CR$ in
powers of the couplings, to derive Schwinger-Dyson equations
which recursively build up planar diagrams from subdiagrams.

\subsec{General solution and derivation of $\CF$}

Since the equation \mmloop\ is quadratic in $\CR$, it is trivial
to write its general solution:
\eqn\loopsoln{
2 \CR(z,\ttheta) =  W'(z) - \sqrt{W'(z)^2 + f(z,\ttheta)} .
}
The sign of the square root is determined by the asymptotics
$\CR(z,\ttheta) \sim 1/z$ at large $z$.

The result determines all expectation values of chiral operators
in terms of the $2n$ bosonic and $2n$ fermionic coefficients of
$f(z,\ttheta)$.  Furthermore, to obtain $W_{eff}$, we need merely
integrate one of these expectation values with respect to its
coupling. Thus we are close to having the solution, if we can
interpret these parameters.

The meaning of the solution is perhaps easier to understand by
considering the component expansion \loopcomponents.  The
$\psi^0$ term is a quadratic equation for $R(z)$, whose solution
is determined by the $n$ coefficients $f_i$ in the
expansion
 $$f(z)=\sum_{i=0}^{n-1} f_i z^i.$$
$R(z)$ is not a single-valued function of $z$, but rather is
single valued on a Riemann surface branched over the $z$ plane;
call this $\Sigma$. Since $W'(z)^2 + f(z)$ is a polynomial of
order $2n$, $\Sigma$  is a Riemann surface of genus $n-1$.

The functions $w_\alpha(z)$ and $T(z)$ are then derived from
$R(z)$ by solving linear equations involving the higher $\psi$
components in $f(z,\ttheta)$.  In particular, the gauge theory
expectation values $$ \bigvev{\Tr \Phi^{n+1}} $$ are encoded in
the expansion of the  component $T(z)$ that is quadratic in
$\psi$, and depend on all of the parameters in $f$.

As we discussed in section 2, there is a related set of $2n+2n$
parameters, the components of $\S_i$ defined in \tindigo,
 \eqn\tindigotwo{\eqalign{ \S_i &= S_i + \psi_\alpha
 w^{\alpha}_i - \psi^1\psi^2 N_i \cr
  &= {1\over 2\pi i}\oint_{C_i} dz \,\CR(z,\ttheta) . }}
These are different from the coefficients in the expansion of
$f(z,\ttheta)$, but the two sets of parameters are simply
related. Substituting \loopsoln\ in \tindigotwo\ and considering
the $\psi^{\alpha} =0$ component, we get
 \eqn\ssg{ S_i = -{1\over 4\pi
 i} \oint_{C_i} dz \sqrt{W'(z)^2 + f(z)} .}  (The polynomial
 $W'(z)$, being non-singular, does not contribute to the contour
 integral.)
We expanded $f(z)$ as $ f(z) = \sum_i f_i z^i . $ Differentiating
with respect to the coefficients $f_i$, one finds
 \eqn\relSf{ {\p S_i \over \p f_j} = - {1\over 8\pi
 i}\oint_{C_i} dz {z^j\over \sqrt{ W'(z)^2 + f(z)}} . }
The integrands in \relSf\ for $0 \le j \le n-2$ provide a complete
set of $n-1$ holomorphic one-forms on the genus $g=n-1$ Riemann
surface $\Sigma$, while the $n$ cycles $C_i$ provide a (once
redundant) basis of A cycles on $\Sigma$. Thus, by taking an
$(n-1)\times (n-1)$ submatrix of \relSf, we have a period matrix,
which for a generic surface $\Sigma$ this matrix will have
non-vanishing determinant.

Finally, by examining the $\CO (1/z)$ term of $R(z)$, the remaining
variable $f_{n-1}$ can be identified with $S=\sum_i S_i$, the
$\psi^0$ component of an integral \tindigotwo\ around a  contour
surrounding all the cuts.  More precisely, one finds
 $$
 R(z) \sim {S\over z} = -{1 \over z} f_{n-1} {n!\over 4
 (\p^{n+1}W/\p z^{n+1})} = -{1 \over z} {f_{n-1}\over 4
 g_n}.
 $$
Combining this relation with \relSf, one finds that the
coordinate transformation $S_i \rightarrow S'_i(S_j)$ is
generically non-singular. Thus, in general either set of
parameters, $S_i$ or $f_j$, could be used to parameterize the
family of surfaces $\Sigma$ which appear in the solutions of
\mmloop.

As we discussed in section 2, the natural fields which specify a
gauge field background in a $\Phi$-independent way are the
$\S_i$.  Thus we can regard the solution \loopsoln\ as a function
of the couplings $g_k$, but at fixed $\S_i$, as giving us the
complete set of expectation values of chiral operators in the
background specified by $\S_i$. We could then obtain the effective
superpotential $W_{eff}$ by integrating any of these expectation
values with respect to its coupling (using \welat), up to a
coupling-independent constant of integration.

Rather than derive $W_{eff}$ directly, we can get the complete
effective action by first deriving the function $\CF_p$ of section
2.  As we discussed there, the $\psi$ shift symmetry tells us that
 $$ W_{eff}= \int d^2\psi\ \CF_p(\S_i,g_k) $$
for some function $\CF_p(\S_i,g_k)$. Using \welat, we have that
 \eqn\upperFdet{\eqalign{ -{1\over 2(k+1)} \int d^2\psi
 \bigvev{ \Tr ({1\over
 4\pi}W_\alpha-\psi_\alpha)^2\Phi^{k+1}}_\Phi &=
 {1\over k+1}\bigvev{\Tr \Phi^{k+1}}_\Phi \cr &=
 \int d^2\psi {\partial \CF_p(\S_i,g_j)\over \partial g_k} .
 }}
This equation does not provide enough information to determine
$\partial \CF_p/\partial g_k$, since there are functions
(depending only on the $\hat S_i$ introduced in \shatsd, for
example) that are annihilated by $\int d^2\psi$. However, all we
are planning to do with $\CF_p$, or its derivative $\partial
\CF_p/\partial g_k$, is to act with $\int d^2\psi d^2\theta$ to
compute the effective action, or its derivative with respect to
$g_k$. As we discussed in introducing \lofog, $\CF_p$ is uniquely
determined if we require it to be a function only of the $\S_i$,
and to vanish at $\S_i=0$.  In that case, $\partial\CF_p/\partial
g_k$ will have the same properties.  The unique $\CF_p$ that has
those properties and obeys \upperFdet\ also satisfies
 \eqn\genwelat{ {\partial \CF_p \over \partial g_k}=-{1\over 2(k+1)}
 \bigvev{ \Tr ({1\over 4\pi}W_\alpha-\psi_\alpha)^2\Phi^{k+1}}_\Phi
 }  By integrating over $\psi$, \genwelat\ clearly implies
 \upperFdet.
\genwelat\ implies the auxiliary conditions on $\CF_p$, since
from section 2, we know that the right hand side of \genwelat\ is
a function only of $\S_i$ and vanishes at $\S_i=0$. That
\upperFdet\ plus the auxiliary conditions on $\CF_p$ imply
\genwelat\ should be clear: no nonconstant function of the $\S_i$
is annihilated by $\int d^2\psi$, so adding such a function to
the right hand side of \genwelat\ would spoil \upperFdet.

Finally, we can complete our argument.  The right hand side of
\genwelat\ is determined as a function of $\S_i$ by \loopsoln. By
integrating with respect to the $g_k$, $\CF_p$ is determined from
\genwelat\ up to a function independent of the $g_k$.  By scaling
the $g_k$, or equivalently scaling the superpotential $W\to
W/\hbar$ with $\hbar$ small, we can reduce the evaluation of
$\CF_p$ to a one-loop contribution, which can be computed
explicitly.  So finally we have shown that the anomalies plus
explicit evaluation of a one-loop contribution suffice to
determine $W_{eff}$.

In summary, using gauge theory arguments, we have derived a
procedure which determines $\CF_p$ up to a $g_k$-independent, but
possibly $S_i$-dependent, contribution.  The subsequent gauge
theory functional integral can also produce such terms, so if we
are discussing the solution of the full gauge theory, the problem
of determining this contribution is not really separable from the
problem of gauge dynamics.

\subsec{Comparison with the Bosonic Matrix Model}

Our arguments so far did not assume any relationship between gauge
theory and matrix models, and did not assume that the gauge theory
reduced to summing planar diagrams.  Indeed, given the
non-perturbative validity of \genkonishian, the arguments are valid
non-perturbatively.

Nevertheless, the final result is most simply described by the
relation to the matrix model.  We very briefly review the matrix
model to explain this point. For a more comprehensive review, see
\mmreview.

\def\hat{\widehat}
The bosonic one matrix model is simply the integral over an $\hat
N \times \hat N$ hermitian matrix $M$, with action $(\hat
N/g_m){\rm tr} \,W(M)$. ($\hat N$ is different from $N$ of the
gauge theory, and will be taken to infinity, while $N$ is held
fixed.)  To make the action dimensionless, we introduced a
dimensionful parameter $g_m$ with dimension $3$ (like $W(\Phi)$).
$W$  in general could be any smooth function with reasonable
properties, but for present purposes, it is the same function
that serves as the superpotential $W(\Phi)$ of the
four-dimensional gauge theory. The free energy $F_{m.m.}$ of the
matrix theory is defined by
 \eqn\mmint{ \exp\left(- {\hat N^2\over g^2_m} F_{m.m.}\right) =
 \int d^{{\hat N^2}} M\ \exp\left(-{{\hat N} \over g_m}\tr
 W(M)\right).
}

This model admits an 't Hooft large $\hat N$ limit, which is
reached by taking  $\hat N\rightarrow\infty$ at fixed $W(M)$ and
$g_m$. As is well known, in this limit, the perturbative expansion
of the free energy $F_{m.m.}$ reduces to planar diagrams.  One
also has many nonperturbative techniques for computing the exact
free energy $F_{m.m.}$, as first explained in \bipz.

One of the many ways to see this is to derive loop equations,
which are Ward identities derived by considering the variations $$
\delta M = \epsilon M^{n+1} $$ or equivalently by using the
identity
 \eqn\kogo{ 0 = \int d^{\hat N^2}M\ \Tr\left({\p\over\p
M}\ M^n\right)
  \exp\left(-{\hat N \over g_m}\tr W(M)\right) .}
   These Ward identities are most simply formulated in terms of the
matrix model resolvent.  For this, one takes a generating
function of the identities \kogo, and writes \eqn\gotto{0 = \int
d^{\hat N^2}M\ \Tr\left({\p\over\p M}\ {1\over z-M} \right)
  \exp\left(-{\hat N \over g_m}\tr W(M)\right) .}
Evaluating this expression, one learns that \eqn\hotto{
\left(g_m\over \hat N\right)^2\left\langle \left(\Tr {1\over
z-M}\right)^2\right\rangle = {g_m\over \hat N} \left\langle \Tr
{W'(M)\over z-M}\right\rangle.} If now we define the matrix model
resolvent,
 \eqn\resolvent{ R_m(z) = {g_m\over
\hat N}\bigvev{\Tr {1\over z-M}} , } then from \hotto, exactly as
in section 3, we deduce that
  \eqn\notto{\left\langle R_m(z)^2\right\rangle =\left\langle
 W'(z)R_m(z)\right\rangle+{1\over 4} f_m(z),}
where $f_m(z)$ is a polynomial of degree $n-1$. Equivalently, we
deduce that
 \eqn\realmmloop{ \left\langle R_m(z)^2 \right\rangle=
 \left\langle\,
 [ W'(z) R_m(z) ]_-\,\right\rangle . }
If now we take $\hat N\to\infty$, then correlation functions
factor in the matrix model,
 \eqn\ealmloop{\left\langle R_m(z)^2\right\rangle =\left\langle
 R_m(z)\right\rangle^2.} In the
gauge theory, the analogous factorization was justified using the
properties of chiral operators, without a large $N$ limit. This
lets us rewrite \notto\ in the form
 \eqn\newnotto{R_m(z)^2= W'(z)R_m(z)+{1\over 4} f_m(z),}
where now the expectation value of $R_m(z)$ is  understood.  Here
we recognize an equation (the first equation in \loopcomponents)
that we have seen in analyzing the gauge theory.

In the matrix model, as explained by Dijkgraaf and Vafa, the
choice of $f_m$ corresponds to the choice of how to distribute
$\hat N$ eigenvalues of the matrix $M$ among the $n$ critical
points of $W$.  Since the gauge theory object $R(z)$ obeys the
same equation  as the matrix model resolvent $R_m(z)$, they will
be equal if $f=f_m$, or equivalently if the fields $S_i$ of the
gauge theory equal the analogous objects in the matrix model.  In
other words, they are equal if
 \eqn\matSi{ S_i = {1\over2\pi i}\oint_{C_i} R_m(z) dz
 ={1\over 2\pi i}\oint_{C_i}{g_m\over \hat N}\bigvev{\Tr{1\over
 z-M}} ={g_m\hat N_i\over \hat N}, }
where $\hat N_i$ is the number of eigenvalues of $M$ near the
$i^{th}$ critical point.  Dijkgraaf and Vafa express this
relation as $S_i=g_s\hat N_i$, where $g_s=g_m/\hat N$.  In the
gauge theory, $R(z)$ depends  on complex variables $S_i$, which
cannot be varied independently if we want them to be of the form
$S_i=g_m\hat N_i/\hat N$. Nevertheless, the holomorphic function
$R(z)$ is determined by its values for $S_i$ that are of this
form, and in this sense the matrix model determines $R(z)$.

The derivative of the matrix model free energy with respect to the
couplings is \eqn\frodo{{\partial F_{m.m.}\over \partial g_k}
=\left\langle {\Tr \,M^{k+1}\over k+1}\right\rangle,} as one
learns by directly differentiating the definition \mmint. So
$R_m(z)$ is the generating function for $\partial F_{m.m.}
/\partial g_k$, just as the gauge theory object $R(z)$ is the
generating function for $\partial \CF/\partial g_k$ at $\psi=0$.
So for gauge theory and matrix model parameters related by \matSi,
we have
 \eqn\mmident{ F_{m.m.}(S_i,g_k)  =
 \CF_p(S_i,g_k)|_{\psi=0} + \CH(S_i),
 }
where $\CH(S_i)$ is independent of $g_k$. This is the relation
between the two models as formulated by Dijkgraaf and Vafa, who
further claim that the left-hand side (or the right hand side
including $\CH$) is the full gauge theory effective
superpotential, including the effects of gauge dynamics, in a
description in which the $S_i$ can be treated as elementary
fields.

Despite the similarities in the gauge theory and matrix model
derivations, there are certainly some differences.  As we have
already noted, the crucial factorization of correlation functions
is justified in the gauge theory using the properties of chiral
operators, and in the matrix model using a large $\hat N$ limit.
Another difference is that in the gauge theory, the variables are
naturally complex, but the parameters and eigenvalues of $M$
naturally take real values. Furthermore, $N_i$ and some other
natural gauge theory quantities have no direct matrix model
analogs.

Now that we have derived the relation to the matrix model, we are
free to use this where it simplifies the arguments, to get
concrete expressions for $\CF(\CS,g)$, and so forth, as we will do
in section 5.

The deepest point where the matrix model produces a stronger
result than our gauge theory considerations so far is in
predicting the coupling-independent factor in $\CF(S,g)$.  It is
easy to compute this, in the matrix model, by taking special
values for the couplings. For example, for the case of unbroken
gauge symmetry, if we take the superpotential $W=m\Phi^2/2$, we
have, following Dijkgraaf and Vafa,
 $$\eqalign{
  \exp\left({-{\hat N^2F_{m.m.}(S,m)\over g_m^2}}\right) &=
 \int {d^{\hat N^2}M \over \mu^{\hat N^2}}\
 \exp\left({-{\hat N m\over 2 g_m}} \Tr M^2\right) \cr
 &= \left({2 \pi g_m\over \hat N m \mu^2}\right)^{\hat N^2/2}}$$
where the constant $\mu$ with dimensions of mass was introduced in
order to keep the measure dimensionless.  From \matSi, $S=g_m$ so
 $$ F_{m.m.}(S,m) = \half S^2 \log \left({m \mu^2\hat N\over 2\pi
 S}\right) . $$
Using \mmident\ and inserting this in \plok, the $m$ dependence
agrees with \gorfun, which is the simplest check of the general
claim.  More impressively, this determines the integration
constant as well.  Identifying $\hat N \mu^2/2\pi$ with the gauge
theory cutoff as $e^{3/2}\Lambda_0^{2}$ (we absorbed the
$1/\Lambda_0^N$ from \gorfun\ here) and again using \plok, we find
precisely \rury.

\subsec{Determining the Full Effective Superpotential}

There are at least two ways one can interpret the final result.
One, which we have emphasized in the introduction as being
conceptually more straightforward, is to consider it as an
effective Lagrangian to be used in the low energy gauge dynamics.
Our arguments seem adequate for proving the result with this
conservative interpretation.

A more ambitious interpretation of the final result is to regard
it as an ingredient in a complete ``effective superpotential,''
obtained by integrating out both matter and gauge fields in a
description in which the $\hat S_i$ are treated as elementary
fields. Despite our caveats in the introduction, one can
certainly make a very simple suggestion for how to determine
this. It is that the Ward identities and derivation we just
discussed apply to the full gauge theory functional
integral.\foot{ As discussed in section 3, the nonrenormalization
arguments so far leave room for $\Lambda$-dependent corrections.
We suggested there that this might not be the case. } After all,
these identities did not explicitly depend on the mysterious
quantities $S_i$ and $w_i$; rather those quantities emerged as
undetermined parameters in the solution.

If we accept this idea, then the remaining ambiguity in the full
effective superpotential is a coupling-independent function of the
variables $S_i$ denoted $\CH (S_i)$ in \mmident.  Since it is
coupling independent, we can compute it in any convenient limit.
In particular, we can consider the limit $W \gg \Lambda_i^3$ in
which the gauge coupling is negligible at the scale of gauge
symmetry breaking.  In this limit, the gauge dynamics reduces to
that of $\prod_i U(N_i)$ super Yang-Mills theory.  And we know
the contribution of this gauge theory to the effective
superpotential -- it is the sum of Veneziano-Yankielowicz
superpotentials \rury\ for each factor
 \eqn\ruryaa{\eqalign{
 &\CH (S_i) = \sum_i N_i H(S_i) \cr
 &{\partial H(S_i) \over \partial S_i} = S_i(1-\ln S_i)}}

Thus, we might conclude that the complete effective superpotential
has the constant of integration fixed
 \eqn\completeWeff{
 W_{eff}(S_i,g_k) = 2\pi i\tau \sum_i S_i + \sum_i N_i  {\p
 \CF(S_i,g_k)\over
 \p S_i}  +\half \sum_{ij}  {\p^2 \CF(S_i,g_k) \over \p S_i \p S_j}
 w_{\alpha i} w^\alpha_j.}
Furthermore, since the shift \tuffo\ is a symmetry of the full
gauge theory, the full effective action must take the form \plok\
with
 \eqn\completeCF{ \CF(S_i) = F_{m.m.}(S_i) , }
with even the coupling independent $S_i$ dependence reproduced,
as conjectured by Dijkgraaf and Vafa.  We believe this conjecture
is correct and the observations we just made are certainly
suggestive.

We point out that since $\CH$ is written as a sum of terms each
depending only on a single $S_i = \widehat S_i -{1\over
2N}w_{\alpha i } w^\alpha_i$, its contribution to the effective
superpotential depends only on the $SU(N_i)$ field $\widehat S_i$.

There are (at least) two more points one would want to understand
to feel that one indeed had a complete proof of the conjecture.
One is to explain why the final step in finding the physical
solution of the gauge theory is to minimize this effective
superpotential. Now on general grounds, if one can convince
oneself that the variables $S_i$ and $w_{i\alpha}$ are fields in
an effective Lagrangian, this would clearly be a correct
procedure.  Now their origin was the parameters determining a
particular solution of the Ward identities \mmloop, and one might
argue that, since one can imagine configurations in which this
choice varies in space and time, the parameters must be fields.

It would be more satisfying to have a better microscopic understanding
of the parameters to justify this argument.  This brings us back to
the other point mentioned in the introduction -- we have no
{\it a priori} argument that the solution should be representable as
an effective Lagrangian depending on such fields at all, and we know
of very similar gauge theory problems in which this is not the case.

So the best we can say with these arguments is that assuming that
the solution of the gauge theory takes the general form \eqform,
we have demonstrated that the fields and effective Lagrangian are
as in \completeWeff.

\subsubsec{Relation to special geometry}

In section 2 we noted that the fields $S_i$, which as we
discussed can be regarded as coordinates on the space of gauge
backgrounds, satisfy many the defining properties of the
``special coordinates'' which appear in $\CN=2$ supersymmetry
(e.g. see \SeibergRS\ for a discussion). In particular they
include gauge field strengths (the $w_{\alpha i}$) normalized so
that the charge quantization condition is independent of $S_i$.

Another observation of \DijkgraafFC\ (stemming from previous work
on geometric engineering of these models) is that the matrix
model free energy $F_{m.m.}(S,g)$, and thus the gauge theory
quantity $F(\CS,g)$, can be regarded as an $\CN=2$ prepotential.
Furthermore, the relations of special geometry can be seen to
follow from matrix model relations.

In particular, the matrix model/loop equation definition \matSi\
of the quantities $S_i$, as periods of non-intersecting cycles
$C_i$ of the Riemann surface $\Sigma$. Special geometry includes
a conjugate relation for the periods of a set of non-intersecting
conjugate cycles $B_j$ whose intersection form satisfies $\langle
C_i,B_j \rangle =\delta_{ij}$. This relation is \eqn\Bperiod{ {\p
F \over \p S_i} = \oint_{B_i} R(z) dz . } In the matrix model,
one can interpret this formula as the energy required to move an
eigenvalue from the $i$'th cut to infinity. This striking
interpretation suggests that the matrix model eigenvalues are
more than just a convenient technical device.  Although the
special geometry relations can be proven in many other ways We do
not know a direct gauge theory argument for this relation.

We end this section by commenting on the situation when some of
the $N_i$ vanish.  In this case the corresponding $S_i$ do not
exist.  This means that the Riemann surface degenerates and the
cycles associated with these $N_i$ are pinched.  In section 5 we
will discuss a simple example in which the Riemann surface
degenerates to a sphere.

\newsec{Examples}

We now grant that supersymmetric vacua of the one adjoint theory
are critical points of the effective superpotential
\completeWeff, and study this solution explicitly.  We consider
two particular cases in which we have other information to
compare with. In the first case the $U(N)$ gauge group is
maximally Higgsed and the low energy gauge group is $U(1)^N$.  In
the second case the $U(N)$ gauge group is maximally confining and
the low energy gauge group is $U(1)$.
Of particular interest is to see how effects usually computed using
instantons can emerge from the procedure of minimizing this effective
superpotential.

\subsec{Maximally Higgsed vacua}

In this subsection we consider the maximally Higgsed vacua in
which the $U(N)$ gauge symmetry is broken to $U(1)^N$. As shown
in \CachazoPR, these vacua are obtained when the tree level
superpotential is of degree $N+1$; i.e.\ $n=N$.  The vacuum with
$\langle\Phi\rangle $ with eigenvalues $\phi_i$ is obtained when
 \eqn\wpri{W'(z)=g_{N}\prod_i(z-\phi_i)}

Since there is no strong dynamics in the IR, we expect all
nonperturbative phenomena to be calculable using instantons.
These are instantons in the broken part of the group and since
they cannot grow in the IR they lead to finite and therefore
meaningful contributions to the functional integral. As the
parameters in the superpotential are varied the low energy
degrees of freedom are the photons of $U(1)^N$ and at some
special points there are also massless monopoles (there can also
be Argyres-Douglas points \refs{\ArgyresJJ, \ArgyresXN}). It is
important that for all values of the parameters the number of
photons remains $N$ and there is no confinement. Therefore, there
is never any strong IR dynamics.

Let us consider the expectation value of $t_k=- {1\over 32\pi^2}
\Tr  W_\alpha^2 \Phi^k$.  Since the light degrees of freedom are
the $N$ photons and at special points also some massless
monopoles, there is no strong dynamics and no IR divergences.
Therefore, there cannot be any singularities as a function of the
parameters in the superpotential $g_k$, and they can be treated
in perturbation theory. Therefore the contribution of $s$
instantons is of the form
 \eqn\generalform{\sum_{r_1,r_2\dots r_n} a_{r_1,r_2\dots r_n}
  \Lambda^{2Ns}g_0^{r_0}g_1^{r_1}\dots g_n^{r_n}}
with coefficients $a_{r_1,r_2\dots r_n}$, which are independent of
the various coupling constants.  Further constraints follow from
the $U(1)_\Phi \times U(1)_\theta$ symmetry of \uonecharges\ and
holomorphy. The $t_k$ have $U(1)_\Phi \times U(1)_\theta$ charges
$(k,2)$.  This leads to the selection rules $2Ns-\sum_l
r_l(l+1)=k$ and $\sum_l 2 r_l= 2$. The only solution for $k\le
N-1$ is $s=1$, $l=N$, $k=N-1$. Therefore,
 \eqn\Tkexp{\langle t_{k}\rangle =\cases{
 0 & $0 \le k \le N-2$ \cr
 c(N) g_N\Lambda^{2N} & $k=N-1$}}
and $\langle t_{N-1} \rangle $ is given exactly by the
contribution of one instanton, and it is first order in $g_N$.
The coefficient $c(N)$ can be computed explicitly, but we will not
do it here. Note that in particular, $\langle S \rangle = \langle
t_0 \rangle =0$.

Let us compare these conclusions with the matrix model. To find
the matrix model curve from the $\N=2$ curve, which is
$y^2=P_N(x)^2-4\Lambda^{2N}$, we are supposed to factor the right
hand side as $P_N(x)^2-4\Lambda^{2N}=Q(x)^2((W'(x))^2+f))$, where
the double roots are contained in $Q(x)^2$.  In the present case,
as the unbroken group has rank $N$, there are no double roots, so
$Q=1/g_N$, $W'=g_NP_N$, and $f=-4g_n^2\Lambda^{2N}$. So
  for $0\le k \le N-1 $, we find
 \eqn\tkexp{\eqalign{
 \langle t_k\rangle =&-{1\over 4\pi i}\oint_\CC z^k  y
 (z) dz= {g_N^2 \Lambda^{2N} \over 2\pi i}\oint_\CC{ z^k \over
 W'(z)} dz \cr
 =& \cases{ 0 & $0 \le k \le N-2$ \cr g_N\Lambda^{2N} &$k=N-1$}}}
which agrees with the expression of $\langle t_k \rangle $ found
in \Tkexp\ for the constant $c(N)=1$.  In deriving \tkexp, we
made the expansion
$y=\sqrt{(W')^2-4\Lambda^{2N}}=W'-2\Lambda^{2N}/W'+\dots$; in
this expansion, only the term $\Lambda^{2N}/W'$ contributes to
the integral.

We mentioned in section 2.2 that also for $N_i=1$ we should still
use $S_i = \widehat S_i - \half w_{\alpha i} w^{\alpha i}$ with a
scalar field $\widehat S_i$ . Classically $\widehat S_i=0$ but
quantum mechanically it can be nonzero. Using contour integrals
we see that $\langle S_i \rangle = \langle \widehat S_i \rangle$
are given by a power series in $\Lambda^{2N}$; i.e.\ they are
given by an instanton sum.  This is a satisfying result because
in this case there is no strong IR dynamics and we do not expect
``gluino condensation in the unbroken $SU(1)$.''  Instead the
nonzero $\langle S_i \rangle$ arise from nonperturbative high
energy effects; i.e.\ instantons.

In appendix A we further study these vacua of maximally broken
gauge symmetry and use our Ward identities to determine the
instanton corrections to $ \langle \Tr  \Phi^k \rangle $ in the
$\CN=2$ theory.  These can be interpreted as a nonperturbative
deformation of the $\CN=2$ chiral ring.

\subsec{Unbroken gauge group}

In this subsection we consider the special vacua in which the
$U(N)$ gauge group is unbroken, and at very low energy only the
photon of $U(1)\subset U(N)$ is massless. These vacua were also
discussed recently in \refs{\FerrariJP,\FujiWD}.

\subsubsec{Strong gauge coupling analysis}

Following \CachazoJY\ we start our discussion by neglecting the
tree level superpotential
 \eqn\supere{W=\sum_{p=0}^{n+1}{g_p\over p+1 }\Tr  \Phi^{p+1}}
Now the theory has $\CN=2$ supersymmetry.  The vacua we are
interested in originate from points in the $\CN=2$ moduli space
with $N-1$ massless monopoles. Turning on the tree level
superpotential \supere\ leads to the condensation of these
monopoles and leaves only a single massless photon.  At these
points in the moduli space the hyper-elliptic curve which
determines the low energy dynamics factorizes as \DouglasNW
 \eqn\ntwofac{y^2=P_N(z)^2-4\Lambda^{2N} =\left((z-z_0)^2-4\Lambda^2
 \right)H_{N-1}^2(z-z_0); \qquad P_N(z)=\det(z-\Phi)}
The constant $z_0$ is present because unlike \DouglasNW, we
consider the gauge group $U(N)$ rather than $SU(N)$ and therefore
the matrix $\Phi$ is not traceless. This factorization problem is
solved by Chebyshev polynomials and determines the locations of
these points in the moduli space of the $\CN=2$ theory at points
where the eigenvalues of $\Phi$ are
 \eqn\maxconpo{\phi_j=z_0+\widehat \phi_j=z_0+2\Lambda
 \cos{\pi(j-\half)\over N}; \qquad\qquad j=1,... ,N}
or related to it by $\Lambda \to e^{2\pi i/2N} \Lambda$ (this
leads to $N$ such points).  Here $\widehat \Phi$ is a traceless
matrix in the adjoint representation of $SU(N)$.  Naively the
expectation value $\langle \Tr \widehat \Phi^k \rangle$ is given
by the sum
 \eqn\sume{\eqalign{
  \sum_{i=1}^N &\left(2\Lambda\cos{\pi(i-\half)\over N} \right)^k
  =\cr
 & N\Lambda^k\cases{0 &$k$ odd\cr
 \pmatrix{k \cr {k\over 2}} & $k$ even, $0 \le k < 2N$ \cr
 \pmatrix{k \cr {k\over 2}} - 2 \pmatrix{k \cr {k\over 2}+N }&
 $k$ even, $ 2N\le k < 4N $\cr
 \pmatrix{k \cr {k\over 2}} - 2 \pmatrix{k \cr {k\over 2}+N }+ 2
 \pmatrix{k \cr {k\over 2}+2N }& $k$ even, $ 4N\le k < 6N $\cr
 }}}
However, in appendix A we argue that $\langle \Tr  \widehat \Phi^k
\rangle$ is corrected by instantons and is not given by the sum
\sume.  Instead
 \eqn\phikch{\langle \Tr \Phi^k \rangle =
 {1\over 2\pi i} \oint_{C} z^k {P_N'(z) \over \sqrt{ P_N(z)^{2}
 -4 \Lambda^{2N}}}dz}
with $P_N$ of \ntwofac, which leads to
 \eqn\hatphin{\langle \Tr  \widehat \Phi^k \rangle =
 \cases{0 &$k$ odd\cr
 N\Lambda^k \pmatrix{k \cr {k\over 2}} & $k$ even}}
for all $k$.

Using \hatphin\ we find an effective superpotential for $z_0$
 \eqn\weffxz{W_{eff}(z_0)=\langle W\rangle =
 N\sum_{l=0}^{\left[{n+1 \over 2}\right]}
 {\Lambda^{2l} \over (l!)^2}  W^{(2l)}(z_0)}

The field $S$ can be integrated in by performing a Legendre
transform with respect to $2N \log \Lambda$. One way to do that
is, as in \tygoa\ by considering
 \eqn\weffxzCS{W_{eff}(z_0,C,S)=N\sum_{l=0}^{\left[{n+1 \over
 2}\right]} {C^{2l} \over (l!)^2} W^{(2l)} (z_0) +
 2 N S\log {\Lambda \over C} }
Integrating $S$ and $C$ out of $W_{eff}(z_0,C,S)$ we recover
$W_{eff}(z_0)$ of \weffxz.  Alternatively, we can integrate out
$C$ to find $W_{eff}(z_0,S)$.  It is obtained from
$W_{eff}(z_0,C,S)$ by substituting $C$ which solves
 \eqn\ceom{0={C\over 2N} \partial_C W_{eff}(z_0,C,S)=
 \sum_{l=0}^{\left[{n+1 \over 2}\right]} {lC^{2l}\over (l!)^2}
 W^{(2l)} (z_0) - S  }
This equation can be interpreted as the Konishi anomaly \konishian\ %
for the rotation of the field $\widehat \Phi$ by a phase.  We can
also integrate out $z_0$ to find $W_{eff}(S)$ by solving its
equation of motion
 \eqn\xzeom{\partial_{z_0} W_{eff}(z_0,C,S)=N\sum_{l=0}^{\left[{n
 \over 2}\right]} {C^{2l} \over (l!)^2}
 W^{(2l+1)}(z_0)=0}
and substituting its solution in $W_{eff}(z_0,C,S)$.

It is interesting that the complexity of $W_{eff}(S)$ as opposed
to the simplicity of $W_{eff}(z_0)$ arises because $S$ was
integrated in and $z_0$ was integrated out.

\subsubsec{Weak gauge coupling analysis}

We now study this vacuum in the limit of weak gauge coupling.
This is the approach we used throughout most of this paper. Rather
than first neglecting the tree level superpotential \supere\ and
adding it later, we first neglect the gauge interactions.  In the
vacuum we are interested in we expand around $\Phi$ which is
proportional to the unit matrix, $\Phi=z_0 $. The traceless
components of $\Phi$ acquire a mass $m(\widehat \Phi)=W''(z_0)$,
and can be integrated out.  We choose not to integrate out $\Tr
\Phi = N z_0$, even though its mass is also $W''(z_0)$. The
remaining light degrees of freedom are the $U(N)$ gauge fields.
Their tree level effective Lagrangian is
 \eqn\treee{\eqalign{
 &{-i\tau \over 16 \pi}\Tr  W_\alpha^2 =
 2\pi i \tau  S\cr
 &\tau ={\theta\over 2\pi}+ i{4\pi \over g_{YM}^2}}}
At one loop order $2\pi i\tau$ is replaced with
 \eqn\lambdai{N\log\Lambda(z_0)^3=N\log\left(\Lambda^2
 W''(z_0)\right)}
which is derived by the threshold matching with the masses of the
massive particles.

Consider the effective Lagrangian at energies above $\Lambda(z_0)$
but below $W''(z_0)$, which is obtained by integrating out the
adjoint massive fields $\widehat \Phi$. The leading order F-terms
are
 \eqn\lowener{NW(z_0)+ 3NS \log \Lambda (z_0)
  + {\rm higher ~ order ~ terms}}
with $\Lambda(z_0)$ of \lambdai.  The higher order terms include
higher powers of $W_\alpha W^\alpha$.  The discussion of section 2
shows that the $F$ terms are only functions of $S$; i.e.\ there
are no terms with a trace of more than two $W_\alpha$, or terms
with different contractions of the Lorentz indices. Furthermore,
terms of order $S^k$ arise only from diagrams with $k$ loops.

Integrating $z_0$ out of \lowener\ leads to more higher order
terms in $S$.  They correspond to diagrams in the high energy
theory which are not 1PI with respect to $z_0$.  Note that these
terms satisfy the nonrenormalization theorem of section 2 and
generate $S^k$ by diagrams with $k$ loops.

As in the rest of the paper we integrate out the strong $SU(N)$
dynamics by adding to \lowener\ the Veneziano-Yankielowicz term
$S(1-\log S)$.  We find the full effective superpotential of
$z_0$ and $S$
 \eqn\lownote{W_{eff}(z_0,S)=N\left[W(z_0)  + S\left[\log \left(
 \Lambda^2 W''(z_0)\over S
 \right)+1\right]+{\rm higher ~ order ~ terms} \right]}

We can compare this expression with the strong gauge coupling
analysis we did earlier in this subsection. If we neglect the
higher order terms in \lownote\ and integrate out $S$, we find
 \eqn\lownoteup{{1\over N} W_{eff}(z_0)= W(z_0)  +
 \Lambda^2 W''(z_0) + \CO(\Lambda^4) }
which agrees with \weffxz.  We can interpret it as follows.  The
first term is due to the tree level superpotential.  The second
term $\Lambda^2 W''(z_0)= \Lambda(z_0)^3$ is due to gluino
condensation in the low energy unbroken group.  Higher order
terms in \weffxz\ can then be interpreted as powers of gluino
condensation.

It follows from \weffxz\ that if the tree level superpotential is
cubic, there are no higher order corrections in \lownoteup, and
therefore there are no higher order corrections in \lownote. This
means that in this case \lownote\ is given exactly by the one
loop approximation.  If we integrate $z_0$ out of \lownote\ to
find an effective Lagrangian for $S$, the term $S^k$ arises from
diagrams with $k$ loops, which are connected by $z_0$ lines and
are not 1PI.  We conclude that in this case of a cubic
superpotential with a maximally confining group only a very small
subset of the $k$ loop diagrams contribute to the term $S^k$.

\subsubsec{Matrix model computations}

In this special case the analysis of the matrix model simplifies
considerably. As mentioned at the end of section 4, when some of
the $N_i$ vanish the Riemann surface degenerates.  In our case of
only one nonzero $N_i$ it degenerates to a sphere. Only one of the
$n$ cuts of the curve
 \eqn\curved{y^2=W'(z)^2 +f(z)}
exists and the other $n-1$ cuts shrink to zero size; i.e.\ $f$
splits only one of the double zeros of $W'(z)^2$ and shifts the
others.  Therefore, equation \curved\ must factorize as
 \eqn\curvedf{y^2=W'(z)^2+f(z)= \left((z-z_0)^2-4C^2
 \right)Q(z)^2}
with $Q$ a polynomial of degree $n-1$.  $z_0$ and $C$ will soon be
identified with the objects denoted by the same symbols above. It
is easy to find the polynomial $Q$ as
 \eqn\findQ{Q(z) = \left[{W'(z) \over \sqrt{(z-z_0)^2-4C^2
 }}\right]_+}
We first solve
 \eqn\curvedfa{W'(z)^2= \left((z-z_0)^2-4C^2
 \right)\widehat Q(z)^2}
for $\widehat Q (z)$ and expand it around $z_0$
 \eqn\hatq{\widehat Q(z)={W'(z) \over\sqrt{(z-z_0)^2-4C^2}}
 =\sum_{k=0}^n \sum_{l=0}^{\infty} { C^{2l} \over k!}
 \pmatrix{2l\cr l} W^{(k+1)}(z_0)(z-z_0)^{k-2l-1}}
where we used ${1\over \sqrt{1-4x}}=\sum_{l=0}^\infty \pmatrix{2l
\cr l}x^l$.  $Q$ in \curvedf\ should be a polynomial and therefore
 \eqn\qpol{Q(z)=\sum_{k=1}^n \sum_{l=0}^{\left[{k-1 \over
 2}\right]} {C^{2l} \over k!} \pmatrix{2l\cr l} W^{(k+1)}(z_0)
 (z-z_0)^{k-2l-1}}
Substituting this $Q$ in \curvedf\ it is easy to see (using
\hatq) that it is satisfied with $f$ a polynomial of degree $n$.
The order $z^n$ term in equation \curvedf\ is cancelled when the
coefficient of $(z-z_0)^{-1}$ in $\widehat Q$ vanishes.
Therefore, equation \curvedf\ can be satisfied only when
 \eqn\condcxz{\sum_{l=0}^{\left[{n\over
 2}\right]} { C^{2l} \over (l!)^2} W^{(2l+1)}(z_0)=0}
Furthermore, the coefficient of $z^{n-1}$ in $f_{n-1}$ is
determined by the coefficient of $(z-z_0)^{-2}$ in $\widehat Q$
 \eqn\Anmo{f_{n-1}=-{4 W^{(n+1)} \over n!} \sum_{l=1}^{\left[
 {n+1\over 2}\right]} {l C^{2l} \over (l!)^2} W^{(2l)}(z_0)}

We can now compare our field theory results with this
discussion.  The relation of $z_0$ and $C$ \condcxz\ is identical
to the equation of motion of $z_0$ \xzeom.  $S$ is determined in
terms of $f_{n-1}$ \CachazoPR
 \eqn\Scurve{S=-{n!\over 4W^{(n+1)}}f_{n-1}=\sum_{l=1}^{\left[
 {n+1\over 2}\right]} {l C^{2l} \over (l!)^2} W^{(2l)}(z_0)}
which is the same as the equation of motion of $C$ \ceom.  This
confirms the identification of $z_0$ and $C$ with the objects
denoted by the same symbols above.

As we reviewed at the end of section 4, the effective
superpotential is determined by the period integral \Bperiod
 \eqn\bitpe{
 -2\pi i \Pi=-\int_{z_0+2C}^{ \Lambda_0} y dz }
In Appendix C we perform this integral and find
 \eqn\biteves{\eqalign{
 -2\pi i \Pi=&-\int_{z_0+2C}^{ \Lambda_0} y dz \cr
 =&-W(\Lambda_0) -2S \log\left({\Lambda_0\over C} \right)
 + \sum_{k=0}^{\left[{n+1\over 2}\right]}
 {C^{2k}\over (k!)^2} W^{(2k)}(z_0) +\CO({1\over \Lambda_0})\cr
 }}
The first term is independent of the fields and can be ignored.
The second and third terms (after renormalizing the bare gauge
coupling) are as in \weffxzCS.  We conclude that as $\Lambda \to
\infty$ we recover the superpotential \weffxzCS\ with $C(z_0)$
determined by \condcxz\ and $z_0(S)$ determined by \Scurve.

Using the explicit expression for $y$ it is straightforward to
compute
 \eqn\tmdefa{\eqalign{
 \langle t_m \rangle =& -{1\over 32 \pi^2} \bigvev{\Tr W_\alpha^2
 \Phi^m}=-{1\over 4\pi i} \oint_\CC z^m y dz=
 -{1\over 4\pi i} \oint_\CC z^m \sqrt{(z-z_0)^2 -4C^2} Q(z) dz \cr
 =& \sum_{k=1 \atop k ~ odd}^n\sum_{r=0}^{\left[{m
 \over 2}\right]} {2 m!\over( 2r+k+1 )
 (r!)^2\left(({k-1\over 2})! \right)^2 (m-2r)!}z_0^{m-2r}
 W^{(k+1)}(z_0) C^{2r+k+1}\cr
 & + \sum_{k=1 \atop k ~even}^n\sum_{r=0} ^{\left[{m-1
 \over 2}\right]} {2 m!\over ( 2r+k+2)({k\over 2})!
 ({k-2\over 2})! r! (r+1)! (m-2r-1)!}\cr
 &\qquad\qquad z_0 ^{m-2r-1}W^{(k+1)}(z_0) C^{2r+k+2}
 }}
Using this expression for $\langle t_m \rangle $ and the
expectation values
 \eqn\vevst{\eqalign{
 &\langle \Tr  \Phi^{k}\rangle = N\sum_{p=0}^{\left[{k\over 2}
 \right] }{k!\over (k-2p)! p!^2}C^{2p}z_0^{k-2p}\cr
 &\langle \Tr  \Phi^{m+1}W'(\Phi ) \rangle = N \sum_{k=0}^{m+1}
 \sum_{l=0 \atop k+l \ {\rm even}}^{n} {1\over l!}
 \pmatrix{k +l\cr (k+l)/2} \pmatrix{m+1 \cr k}
 z_0^{m+1-k} W^{(l+1)}(z_0) C^{k+l}}}
it is straightforward to explicitly check the generalized anomaly
identity
 \eqn\relat{\langle \Tr  \Phi^{m+1}W'(\Phi ) \rangle  = 2
 \sum_{l=0}^m \langle \Tr  \Phi^{m-l} \rangle \langle t_l \rangle}
or equivalently
 \eqn\relats{\bigvev{ \Tr {W'(\Phi ) \over z-\Phi}}  = 2
 \bigvev{ \Tr {1 \over z-\Phi} }\bigvev{-{1\over 32 \pi^2}
  \Tr { W_\alpha^2 \over z-\Phi}}.}

Instead of checking \relat\ or \relats\ explicitly, we can
proceed as follows.  We use the matrix model loop equation
 \eqn\lopp{ \left( -{1\over 4\pi i} \oint_{\CC} { y(z)\over z-x}
 dz \right)^2 = -{1\over 4\pi i} \oint_{\CC }
 {W'(z) y(z)\over z-x} dz }
where $\CC$ is a large circle at infinity and our identification
 \eqn\walphasaa{-{1\over 32 \pi^2}\bigvev {\Tr {W_\alpha^2
 \over z-\Phi}} = - {1\over 4\pi i} \oint_{\CC} {y(x)\over z-x} dx}
to show that the values obtained in the field theory analysis for
$\langle \Tr \Phi^k \rangle$ \vevst\ or $\bigvev{\Tr {1\over
z-\Phi}}$ satisfy \relats.

Equations \vevst\ can be written as\foot{From differentiating
\ntwofac, $H_{N-1}$ divides $P_NP'_N$. However, \ntwofac\ shows
that $P_N$ and $H_{N-1}$ cannot have the same root and therefore,
$P_N'$ has a factor of $H_{N-1}$.  Since they are polynomials of
the same degree, they are proportional to each other.  The
proportionality factor can be determined by matching the highest
power in the polynomial, and therefore $P_N'=NH_{N-1}$.  Finally,
the equations in appendix A show that we need ${P_N' \over
\sqrt{P_N^2 -4 \Lambda^{2N}}}= {N\over \sqrt{(z-z_0)^2-4C^2}}$.}
 \eqn\trphizaa{\eqalign{
 &\langle \Tr \Phi^k   \rangle = {N\over 2\pi i} \oint_{\CC}
 {z^k\over  \sqrt{(z-z_0)^2-4C^2}} dz \cr
  &\langle \Tr \Phi^k W'(\Phi)   \rangle = {N\over 2\pi i}
  \oint_{\CC} {z^k W'(z)\over  \sqrt{(z-z_0)^2-4C^2}} dz \cr}}
or equivalently
 \eqn\trphiza{\eqalign{
 & \bigvev {\Tr {1 \over z-\Phi}} =  {N\over 2\pi i}
 \oint_{\CC} {1\over (z-x)\sqrt{(x-z_0)^2-4C^2}} dx \cr
  & \bigvev {\Tr {W'(\Phi)\over z-\Phi}} =  {N\over 2\pi i}
 \oint_{\CC} {W'(x) \over (z-x)\sqrt{(x-z_0)^2-4C^2}} dx
 }}
In appendix C we show that for the $y$ of \curvedf,
\eqn\jkk{{\p y\over \p S} = -  {2\over \sqrt{(z-z_0)^2-4C^2}}. }
and therefore \trphizaa\ can be written as
 \eqn\trphizc{ \eqalign{
 &\bigvev {\Tr {1 \over z-\Phi}} = - {N\over \pi i}
 {\p \over \p S}\oint_{\CC} {y(x)\over z-x} dx \cr
 &\bigvev {\Tr {W'(\Phi) \over z-\Phi}} = - {N\over \pi i}
 {\p \over \p S}\oint_{\CC} {W'(x) y(x)\over z-x} dx \cr
 }}
The derivative of \lopp\ with respect to $S$ is
 \eqn\loopS{  2\left({1\over 4\pi i} {\p \over \p S}  \oint_{\CC}
 { y(z)\over z-x} dz \right) {1\over 4\pi i} \oint_{\CC} {y(z)\over
 z-x}  dz   = -{1\over 4\pi i} {\p \over \p S} \oint_{\CC} {W'(z)
  y(z)\over z-x} dz }
Now equations \walphasaa\trphizc\ leads to the desired result
\relats.

\newsec{Conclusions}

Following the ideas of Dijkgraaf and Vafa, and specifically their
observation that the superpotential in $\CN=1$ gauge theory arises
entirely from the planar diagrams of an associated bosonic matrix
model, we were able to derive a complete solution for the general
$\CN=1$ gauge theory with a single adjoint chiral multiplet,
confirming their conjectured solution.

Besides confirming their results, we have provided an explanation
for the conjecture, which allowed us to make a nonperturbative
proof, and see clearly how far it should generalize. The
reduction to planar diagrams is analogous to the explanation of
the reduction of perturbation theory to planar diagrams in the 't
Hooft large $N$ limit, but it is essentially different.  In
particular it happens for finite $N$.

We can identify four essential ingredients in this proof: the use
of the complete chiral ring, the generalized Konishi anomaly,
factorization for vacuum expectation values of chiral operators,
and the correct gauge-invariant identification of the gaugino
condensates. Let us discuss these points in turn.

First of all, we made essential use of the complete chiral ring
in the gauge theory, which we showed in the single adjoint theory
consists of $\Tr \Phi^n W_\alpha^k$ for all $n$ and all $k\le
2$.  This explicit description of the chiral ring shows up in
some previous work (for example on the AdS/CFT correspondence
\agmoo) but was not systematically exploited in solving the
theories.

One reason for this is that in $U(N)$ gauge theory at any finite
$N$, one expects that all of the operators with $k>N$ can be
written in terms of a finite $\CO(N)$ subset of them, using
operator relations which are simple deformations of those in the
classical theory.  We illustrated this point and its power by
using it to show that in pure super Yang-Mills theory, every
vacuum has a non-zero gaugino condensate.  One might think that
knowing these finite $N$ relations would be essential in solving
the theory, but in the end we were able to do it using only a
subset of relations which survive the large $N$ limit.

The description of the chiral ring is simplified by making use of
a nonlinearly realized $\CN=2$ supersymmetry, which is simply the
shift of the gaugino for the decoupled diagonal $U(1)$ subgroup
of the gauge group.  This symmetry leads to many parallels
between the subsequent discussion and the discussion of gauge
theory with $\CN=2$ supersymmetry.  On first sight, these
parallels might have led one to think that the Dijkgraaf-Vafa
conjectures would only apply to theories (such as our example)
with an explicitly broken $\CN=2$ supersymmetry. However, this
particular nonlinearly realized $\CN=2$ is present in any gauge
theory with a decoupled $U(1)$.  Indeed, our proof of the
conjecture applies to a very large class of $\CN=1$ gauge
theories, including chiral theories.

Our second ingredient was a generalization of the Konishi anomaly,
an anomaly in the symmetry $\delta \Phi = \epsilon \Phi$, to the
general reparameterization of the chiral ring $\delta \Phi =
f(\Phi,W_\alpha)$. The analogous algebra of reparameterizations
of a bosonic matrix integral leads to the Virasoro constraints
and is thus central both to many methods for its solution and to
its connection to topological string theory. In gauge theory,
these considerations lead to a $\CN=2$ super-Virasoro algebra of
constraints\foot{ Making this point explicit requires introducing
couplings for all of the chiral operators.  We did not need this
for the rest of our arguments, and in fact giving finite values
to these couplings is problematic because then the kinetic term
of the scalar fields is not positive definite, but it is a nice
formal way to understand the constraints.}.

Perhaps the simplest consequence of the generalized Konishi
anomaly is a closed set of equations for expectation values of the
``generalized gaugino condensate,'' the operator $\Tr
W_\alpha^2/(z-\Phi)$.  This leads to the identification of this
operator with the matrix model resolvent.

Our third ingredient was the factorization of gauge-invariant
expectation values for chiral operators.  This is perhaps the most
striking point where the discussion for the matrix model and the
gauge theory, while formally parallel, in fact rest on rather
different physics.

Let us explain this point.  In both cases, the key observation is
factorization of correlation functions of gauge-invariant
operators. This allows rewriting the Schwinger-Dyson equations,
which generally relate $n$-point functions to $n+1$-point
functions, as a closed system of equations for the one-point
functions.  The resulting equations can be expanded in the
couplings and seen to generate planar perturbation theory.
However, if their derivation is valid nonperturbatively, then so
are the equations.  In this case, one wants a more conceptual
description of the simplification which does not rely on
planarity of Feynman diagrams for its formulation.

In the large $N$ limit, the deeper reason for factorization is the
existence of a ``master field,'' a single field configuration
which dominates the functional integral in this limit.  It can be
determined by solving a corrected ``equation of motion,'' derived
from an effective action which is the sum of the classical action
with a ``quantum entropy'' term which is the logarithm of the
volume of the gauge orbit of a configuration.  Both terms scale
as $\CO(N^2)$ and thus one can think of the master field as a
saddle point dominating the functional integral.  This
description is nonperturbative and can be used to exhibit many
phenomena (such as certain large $N$ transitions) which have no
diagrammatic interpretation.

In supersymmetric gauge theory, the deeper reason for
factorization is the position independence of chiral operators
combined with cluster decomposition.  There is no obvious sense
in which a single field configuration dominates the functional
integral.  Instead, the correlation function of chiral operators
is given by a single state, the vacuum, in all possible channels.
Yet the formal conclusions are the same, a system of factorized
Ward identities which determine all expectation values of chiral
operators.

Finally, the last ingredient in the proof was to identify the
undetermined parameters in the general solution of the Ward
identities, expressed in a correct set of coordinates, as fields
in an effective Lagrangian.  This is one of the points at which
previous attempts to understand generalizations of the
Veneziano-Yankielowicz effective Lagrangian ran into difficulties,
so let us also explain this point further.

It was long realized that in simple limits (e.g.\ when the
underlying gauge symmetry is broken to a product of nonabelian
subgroups at very high energy and thus very weak couping), the
physics of the theory under discussion would be well described by
a sum of a Veneziano-Yankielowicz effective superpotential in
each unbroken subgroup, each depending on a gaugino condensate
for its subgroup. The problem with making sense of this
observation more generally is that there is no obvious way to
make this definition gauge invariant, as quantum fluctuations do
not preserve the simple picture of subgroups sitting in the
complete gauge group in a fixed way.  Because of this, it is not
{\it a priori} obvious that these variables have any meaning in
the strong coupling regime.

The only gauge invariant way to pick out subgroups of the overall
gauge group is to use the expectation values of the matter fields
to do it.  So, one can try to write an operator such as ``$\Tr
W_\alpha^2 \delta(\Phi - a_i)$'' to pick out the gaugino
condensate associated to the vacuum $W'(a_i)=0$.  However, this
definition would also appear to behave in a very complicated way
under adding quantum fluctuations.

The resolution of this problem is rather striking.  Because of
holomorphy, one can replace the $\delta(\Phi-a_i)$ of the
previous definition with a contour integral $\oint dz/(z-\Phi)$
where the contour surrounds the classical vacuum $a_i$.  Note
that this object involves all powers of $\Phi$.  This is
manifestly independent of infinitesimal quantum fluctuations of
$\Phi$ and thus provides a good definition to all orders in
perturbation theory.  However it is not {\it a priori} obvious
that it provides a good nonperturbative definition.

Of course, in the explicit solution, these poles turn into cuts at
finite coupling, so by enlarging the contours to surround the cuts
one does obtain a good nonperturbative definition.  However this
point could not have been taken for granted but actually rests on
the previous discovery that computations in the chiral ring
reduce to planar diagrams.  Let us explain this point.

An important general feature of the sum over planar diagrams is
that, while the total number of diagrams at a given order $V$ in
the couplings $g$ (and thus weighed by $g^V$) grows as $V!$, the
number of planar diagrams only grows exponentially in $V$.
Furthermore, since we are integrating out massive fields and
obtaining cutoff-independent results each diagram makes a finite
contribution.  This means that, in contrast to the typical
behavior of perturbation theory, the perturbative expansion for
the effective matter superpotential has a finite radius of
convergence.

This makes it sensible to claim that this expansion converges to
the exact result.  Holomorphy precludes nonperturbative
corrections because they have an essential singularity at the
origin. Therefore the expansion must converge to the exact result.

Thus, perturbation theory is much more powerful than one might
have expected, because of the reduction to planar diagrams
combined with holomorphy.  Indeed, the most striking aspect of
the Dijkgraaf-Vafa conjecture is the possibility to derive
effective actions which from other points of view involve
infinite sums over instanton corrections, without any need to
explicitly discuss instantons.  On the other hand, even if one
can derive correct results by summing perturbation theory, it is
very useful to have a nondiagrammatic framework and derivation to
properly justify and understand the results.

There are a number of fairly clear questions coming out of this
work which we believe could be answered in the near future.  A
particularly important question is to better understand the chiral
ring and the relations which hold at finite $N$.  We discussed the
exact quantum relations in some cases; having a simple and
general description would significantly clarify the theory.

Our arguments generalize straightforwardly to arbitrary $U(N)$
quiver theories, with the inhomogeneous $\CN=2$ supersymmetry,
loop equations, and so forth, including the general conjectured
relation between these theories and the corresponding
multi-matrix models. Although the list of matrix models we know
how to solve exactly is rather limited and will probably stay
that way, one can of course get perturbative results.  There is
also a general mathematical framework called free probability
theory \voiculescu\ which allows discussing these models
nonperturbatively and may eventually be useful in this context.

It will be interesting to generalize these arguments to the other
classical gauge groups.  Since we do not rely on large $N$, it
may be that even exceptional groups can be treated.

Although we feel we have provided a more precise connection
between supersymmetric gauge theory and matrix models than given
in previous work, we have by no means understood all of the
connections or even all of the aspects of Dijkgraaf and Vafa's
work, from the point of view of gauge theory.  We mentioned two
particularly striking points in section 4; the fact that the
matrix model naturally reproduces the complete effective
superpotential coming from gauge theory, and the deeper aspects
of the connection to $\CN=2$ supersymmetry and special geometry.
We believe the first point will obtain a purely gauge theoretic
explanation, perhaps by considering additional anomalies.

Regarding the second point, given the way in which the solution
for the adjoint theory is expressed using periods and a Riemann
surface, one can make a simple proof of the relation to special
geometry (expressed for example in \Bperiod), but in itself this
is not a very physical argument.  A more physical argument for
this relation in this example uses the relation to $\CN=2$ super
Yang-Mills and holomorphy.  If this were the only correct
argument, the relation might not be expected to hold in more
general $\CN=1$ theories.  On the other hand, clearly many more
$\CN=1$ theories, including chiral theories, can be obtained by
wrapping branes in Calabi-Yau compactification of string theory,
a situation in which one might think special geometry would
appear, so it could be that there are more general arguments.

Perhaps the largest question of which this question is a small
part is to what extent this structure can be generalized to
describe all of the $\CN=1$ supersymmetric compactifications of
string theory. Although the present rate of progress may be
grounds for optimism, clearly we have much farther to go.

\centerline{\bf Acknowledgements}

This work was supported in part by DOE grant \#DE-FG02-96ER40959
to Rutgers, and DOE grant \#DE-FG02-90ER40542 and NSF grant
\#NSF-PHY-0070928 to IAS.

\appendix{A}{Comments on the Chiral Ring of Pure Gauge $\CN=2$
Theories}

Following \CachazoPR\ in this appendix we learn about the $\CN=2$
theory by first perturbing it by a superpotential and then turning
off the perturbation.

Let us recall some well known facts about the classical chiral
ring of $\CN=2$ theories. The Coulomb branch of the $U(N)$ gauge
theory with $\CN=2$ supersymmetry can be parametrized by the
eigenvalues $\phi_i$ of the $N\times N$ matrix $\Phi_{cl}$
(modulo permutations). Equivalently, in terms of the quantum
field $\Phi$ we can take
 \eqn\modulic{\langle u_k \rangle = \langle \Tr \Phi^k \rangle =
 \Tr  \Phi_{cl}^k =\sum_{i=1}^N \phi_i^k; \qquad\qquad k=1,...,N.}
as good coordinates on the Coulomb branch.

Classically it is clear that $u_k$ generate the chiral ring of
this theory.  To see that, recall that $\Phi_{cl}$ is an $N\times
N$ matrix and therefore for $l>N$ we can express the classical
elements of the ring $\Tr \Phi_{cl}^l $ as polynomials in the
classical generators $\Tr \Phi_{cl}^k$ with $k=1,...,N$
 \eqn\classrelnt{\Tr  \Phi_{cl}^l = \CP_l(\Tr  \Phi_{cl}, \Tr
 \Phi_{cl}^2,...,\Tr  \Phi_{cl}^N)}
These relations can be nicely summarized by introducing the
characteristic polynomial of $\Phi_{cl}$, $P_N(z,\Phi_{cl}) =\det\
(z-\Phi_{cl})$, which can be expressed as an explicit polynomial
in $\Tr\,\Phi_{cl}^k$, $k\leq N$.   In order not to clutter the
equations, we will suppress the second argument $\Phi_{cl}$ in
$P_N$.  Then,
 \eqn\classrel{ \Tr  {1\over z-\Phi_{cl}}={P_N'(z)\over P_N(z)}.}
The desired relations can be found by expanding \classrel\ in
powers of $1/z$ or equivalently
 \eqn\trpka{\Tr  \Phi_{cl}^l ={1\over 2\pi i} \oint_{\CC}
 z^l \Tr {1\over z-\Phi_{cl} }  dz =\oint_{\CC}
 z^l {P_N'(z)\over P_N(z)} dz}
where $\CC$ is a large contour around $z=\infty$.

We expect the relations \classrelnt\ to be modified by instantons.
In order to determine these modifications we follow \CachazoPR\
and deform the $\CN = 2$ theory to $\CN=1$ by adding a
superpotential of degree $N+1$, i.e. $n=N$.  We take $W'(x) = g_N
P_N(x)$ with $P_N(x)=\det(x-\Phi_{cl})$. As in section 5.1, we
expand around the vacuum where the $U(N)$ group is broken to
$U(1)^N$.  This vacuum is uniquely characterized by $\langle u_k
\rangle$ expressed in terms of $\Phi_{cl}$ as in \modulic.

We can solve the theory, as in section 4, using the Ward
identities \loopcomponents\ (with $w_\alpha$ set to zero as we
will be considering expectation values)
 \eqn\wardidnta{\eqalign{ &R(z)^2 =  g_N P_N(z) R(z)
 + {1\over 4}f(z) \cr
 &2 T(z) R(z) =  g_N P_N(z) T(z) + {1\over 4}c(z). }}
The solution of these equations is
 \eqn\solvwa{\eqalign{
 &R(z) ={1\over 2}\left( g_N P_N(z) - \sqrt{ g_N^2 P_N^2(z) +
  f(z) } \right) \cr
 &T(z) = - {c(z) \over  4\sqrt{ g_N^2 P_N^2(z) + f(z) }}.}}
The polynomial $c(z)$ depends only on $\langle \Tr \Phi^k \rangle
= \sum_i \phi_i^k $ for $k=1,...,N-1$
 \eqn\polce{c(z)= 4\left\langle \Tr {W'(\Phi) - W'(z) \over
 z - \Phi} \right\rangle = - 4W''(z)= - 4g_NP_N'(z)}
In a vacuum with unbroken $U(1)^N$ and $W'=g_NP_N$,
$f(z)=-4g_N^2\Lambda^{2N}$, as we explained in deriving \tkexp.
Therefore, the quantum modified version of \classrel\ is
 \eqn\quantrel{ T(z) = \left\langle \Tr  {1\over z-\Phi
 } \right\rangle= {P'_N(z) \over  \sqrt{ P_N^2(z)
 -4\Lambda^{2N} }}.}
Notice that the $g_N$ dependence drops out of the equation and
therefore we can take it to zero.  We conclude that \quantrel\ is
satisfied in the pure $\CN =2$ theory.

Using \quantrel\ we can write,
\eqn\phiqu{\langle \Tr \Phi^l \rangle={1\over 2\pi i} \oint_{C}
z^l {P_N'(z) \over  \sqrt{ P_N^2(z)
 -4\Lambda^{2N} }}dz = \sum_{m=0}^{\left[{2N \over l}\right]}
\pmatrix{2m\cr m} \Lambda^{2 N m}  {1\over 2\pi i} \oint_{C} z^l
{P_N'(z) \over P_N(z)^{2m+1}}dz}
The $m=0$ term is the classical formula \trpka, and the other
terms are generated by $m$ instantons.  This equation was
conjectured in \DijkgraafPP\ and has been used and explored in
\refs{\GopakumarWX,\Schnitzer}.

Once we establish this expectation value we can derive the
modifications of the relations in the chiral ring.  We replace
\classrelnt\ with
 \eqn\quanrelnt{\Tr  \Phi^l = \CQ_l(\Tr  \Phi, \Tr
 \Phi^2,...,\Tr  \Phi^N, \Lambda^{2N})}
Let us compute the expectation value of this equation
 \eqn\quanrelntf{\langle \Tr  \Phi^l \rangle = \CQ_l(\langle\Tr
 \Phi \rangle, \langle\Tr
 \Phi^2\rangle,...,\langle\Tr  \Phi^N\rangle, \Lambda^{2N}) =
 \CQ_l(\Tr \Phi_{cl} , \Tr \Phi_{cl} ^2,...,\Tr \Phi_{cl} ^N,
 \Lambda^{2N}) }
Comparing with \phiqu\ we find the polynomials $\CQ_l$.  The
important point is that once we establish that the ring is
modified as in \quanrelnt, we can use this equation not only in
expectation values of the one point function but also in all
correlation functions.

Before closing this appendix, we comment on the $\CN=1$ chiral
operators $\Tr  W_\alpha \Phi^k$ and $\Tr  W_\alpha W^\alpha
\Phi^k$.  They are related by the $\CN=2$ algebra to the chiral
operators $\Tr\,\Phi^k$.  Hence, although they are $\CN=1$ chiral,
they are not chiral with respect to the $\CN=2$ symmetry.
Therefore, they play no role in the chiral ring of $\CN=2$
theories.

\appendix{B}{The Classical Chiral Ring in $SU(N)$ with $\CN=1$}

In this appendix, filling in some details from section 2.2, we
establish the classical ring relation $S^N=0$ for $SU(N)$
gluodynamics, and we show that $S^{N-1}\not= 0$.

We start with the following identity, which holds for any
$N\times N$ matrix $M$:
 \eqn\prelation{\det M = {1\over N!}(\Tr M)^N + \dots.}
Here the ellipses are a sum of products of traces of $M$, with
 each term in the sum being proportional to at least one factor of
$\Tr M^k$ with some $k>1$.  (To prove this relation, note that if
$\lambda_1,\dots,\lambda_N$ are the eigenvalues of $M$, then
$\det M=\lambda_1\lambda_2\cdots\lambda_N$, while $(\Tr M)^N
=N!\lambda_1\lambda_2\cdots\lambda_N+\dots$, the ellipses being a
sum of terms proportional to some $\lambda_j^k$ with $k>1$. Those
latter terms contribute the terms represented by ellipses in
\prelation.) Now, set $M=W_1W_2$. The ellipses in \prelation\ are
non-chiral terms, and $(\Tr M)^N$ is a multiple of $S^N$, so if
we can show that $\det M$ is non-chiral, it follows that $S^N$
vanishes in the chiral ring. To show that $\det M$ is non-chiral,
we write first \eqn\longfirst{\det M =
(-1)^{N(N-1)/2}\epsilon_{i_1i_2\dots i_N}\epsilon^{k_1k_2\dots
k_N}W^{i_1}{}_{j_1}W^{i_2}{}_{j_2}\dots W^{i_N}{}_{j_N}\tilde
W^{j_1}{}_{k_1}\tilde W^{j_2}{}_{k_2}\dots \tilde
W^{j_N}{}_{k_N},} where to minimize the number of indices, we
write $W$ and $\tilde W$ for $W_1$ and $W_2$. We will focus on the
factor \eqn\kuglo{\epsilon_{i_1i_2\dots
i_N}W^{i_1}{}_{j_1}W^{i_2}{}_{j_2}\dots W^{i_N}{}_{j_N}.} This is
symmetric in $j_1,\dots,j_N$, so we may as well set those indices
to a common value, say $N$, replacing \kuglo\ with
\eqn\kglo{\epsilon_{i_1i_2\dots
i_N}W^{i_1}{}_{N}W^{i_2}{}_{N}\dots W^{i_N}{}_{N}.}
 We will show
that \kglo\ is a multiple of the non-chiral quantity
\eqn\gfirst{\eqalign{\epsilon_{i_1i_2\dots i_{N-1} N}&
W^{i_1}{}_{N}W^{i_2}{}_{N}\dots W^{i_{N-2}}{}_{{N}} \left\{\bar
Q^{\dot\alpha },[ D_{1,\dot\alpha},W^{i_{N-1}}{}_{N}]\right\}\cr
& =\epsilon_{i_1i_2\dots i_{N-1} N}
W^{i_1}{}_{N}W^{i_2}{}_{N}\dots
W^{i_{N-2}}{}_{N}\{W,W\}^{i_{N-1}}{}_{N}\cr &
=\epsilon_{i_1i_2\dots i_{N-1} N} W^{i_1}{}_{N}W^{i_1}{}_{N}\dots
 W^{i_{N-2}}{}_{N}W^{i_{N-1}}{}_xW^x{}_{N}\cr}}
where $ D_{1,\dot \alpha}$ is the bosonic covariant derivative
$D/Dx^{\alpha\dot\alpha}$ with $\alpha=1$. Our goal is now to
show that this quantity is a non-zero multiple of the quantity in
\kglo. If we set $x=N$, we do get such a multiple. If we set $x$
to be one of $i_{1},\dots,i_{N-2}$, we get an expression that
vanishes by Fermi statistics.  And finally, if we set
$x=i_{N-1}$, we get $W^{i_{N-1}}{}_{i_{N-1}}=\Tr\, W=0$, since
$W$ takes values in the Lie algebra of $SU(N)$. A more precise
way to express this argument is to use Fermi statistics to write
\eqn\toxpress{ W^{i_1}{}_NW^{i_2}{}_N\cdots W^{i_{N-2}}_N
W^x{}_N={1\over (N-1)!}\epsilon^{i_1\dots
i_{N-2}xy}\epsilon_{m_1m_2\dots m_{N-1}y}
W^{m_1}{}_NW^{m_2}{}_N\dots W^{m_{N-1}}{}_N.} Inserting this in
\gfirst, one then expresses $\epsilon_{i_1i_2\dots i_{N-1}
N}\epsilon^{i_1\dots i_{N-2}xy}$ as a multiple of
$\delta^x_{i_{N-1}}\delta^y_N-\delta^y_{i_{N-1}}\delta^x_N$. The
first term then gives a multiple of $\Tr\,W=0$, and the second
gives the desired multiple of \kglo.

To complete the picture, we should also prove that $S^{N-1}\not=
0$ in the chiral ring; in other words, we should prove that for
any $X^{\dot\alpha}$, \eqn\loopy{S^{N-1}\not= \{\bar Q_{\dot
\alpha},X^{\dot\alpha}\}.} Any expression of the form $\{\bar
Q_{\dot\alpha},X^{\dot\alpha}\}$ that can be written as a
polynomial in  $W_\alpha$ is proportional to an anticommutator
$\{W_\alpha,W_\beta\}$, since the action of $\bar Q_{\dot\alpha}$
on a bosonic covariant derivative generates
$\{W_\alpha,\cdots\}$, and acting on anything else there is no
way to generate an expression involving only $W_\alpha$.
(Conversely, we have seen that any polynomial in $W$ proportional
to $\{W_\alpha,W_\beta\}$ is of the form $\{\bar
Q_{\dot\alpha},X^{\dot\alpha}\}$.) If, therefore, we can find a
gauge field configuration in which $\{W_\alpha,W_\beta\}=0$ for
all $\alpha,\beta$, and in which $S^{N-1}\not= 0$, this will
establish that $S^{N-1}\not=\{\bar
Q_{\dot\alpha},X^{\dot\alpha}\}$ for any $X^{\dot\alpha}$.  We
simply take an abelian background connection for which $W_\alpha$
is diagonal, $W_\alpha={\rm
diag}(\psi_\alpha^1,\psi_\alpha^2,\dots,\psi^N_\alpha)$; here
$\psi_\alpha^i$ are anticommuting $c$-numbers obeying
\eqn\jurry{\sum_i\psi_\alpha^i=0,} but otherwise unconstrained. So
$S=-{1\over 32\pi^2} \sum_{i=1}^N\psi_\alpha^i\psi^{\alpha\,i}$,
from which it follows that for this configuration $S^{N-1}\not=0$,
though \jurry\ implies that this configuration has $S^N=0$.

\appendix{C}{Computing the Period Integral}

The purpose of this appendix is to compute the period integral in
the case of a maximally confining gauge group. We need to
calculate
 \eqn\intdf{ \eqalign{
 -2\pi i \Pi(z_0,\Lambda_0 ) =& -\int_{z_0+2C}^{\Lambda_0} y dz
 \cr
 y^2=&W'(z)^2+f(z)= \left((z-z_0)^2-4C^2 \right)Q(z)^2}}
with $C$ a solution of
 \eqn\condd{ \sum_{l=0}^{\left[{n\over
 2}\right]} { C^{2l} \over (l!)^2} W^{(2l+1)}(z_0)=0}
and $Q$ given by \qpol.

We are interested in the limit of $\Pi(z_0, \Lambda_0)$ as
$\Lambda_0 \to \infty$. The asymptotic behavior of $y$ is
\eqn\assy{ y = W'(z) +\half {f_{n-1} n! \over W^{(n+1)}} {1\over
z} + \CO({1\over z^2}) }
with
\eqn\fncc{ f_{n-1} = - 4 {W^{(n+1)}\over n!} S =  - 4 {W^{(n+1)}
\over n!} \sum_{l=1}^{\left[n+1\over 2 \right]} {l C^{2l}\over
(l!)^2}W^{(2l)}(z_0). }
Therefore by integrating \assy,
\eqn\piec{-2\pi i \Pi (z_0,\Lambda_0 ) = -W(\Lambda_0) + 2 S \log
\Lambda_0 + F (z_0)+ \CO \left( {1\over \Lambda_0}\right). }
We should determine the $\Lambda_0$ independent function $F(z_0)$.

Let us differentiate \piec\ with respect to $S$. In doing that we
should remember that $z_0$ is a function of $S$ through the
relation \fncc.
\eqn\comp{2 t(S)=-2\pi i{\p \over \p S} \Pi(z_0,\Lambda_0 ) =  2
\log \Lambda_0 + {\p \over \p S} F(z_0) +  \CO \left( {1\over
\Lambda_0}\right)  }
Although we will not need it in the derivation, we comment that
this $t(S)$ is the coupling denoted by $t(S)$ in \newform.

We can also compute $2t(S)=-2\pi i {\p \over \p S}
\Pi(z_0,\Lambda_0 )$ from the integral expression,
\eqn\easy{ 2t(S)=-2\pi i {\p \over \p S} \Pi(z_0,\Lambda_0 ) =  -
\int_{z_0+2C} ^{\Lambda_0}{\p \over \p S} y dz }
where the term from the derivative acting on the lower bound of
the integral is zero since $y(z_0+2C) = 0$ as can be seen in
\intdf.

Let us determine the meromorphic function ${\p \over \p S} y(z)$
by its behavior at its singular points.
 $$ {\p y\over \p S} \to  \cases{
 - {2\over z} & for $z\to \infty$ \cr
 {c_1\over \sqrt{z-z_0+2C}} & for $z\to z_0-2C$ \cr
 {c_2\over\sqrt{z-z_0-2C}} & for $z\to z_0+2C$. } $$
with $c_i$ two non zero constants. The unique meromorphic
function with this behavior is,
$$  {\p y\over \p S} = -  {2\over \sqrt{(z-z_0)^2-4C^2}} $$
The integral \easy\ is elementary and is given by,
$$  2t(S)= 2 \log \left( {\Lambda_0 - z_0 + \sqrt{(\Lambda_0
-z_0)^2-4C^2} \over 2C}  \right) =  2 \log{\Lambda_0\over C} +
\CO \left( {1\over \Lambda_0}\right) $$
Comparing this result with \comp\ we get,
\eqn\codo{ {\p \over \p S} F (z_0) =- 2 \log{C(z_0)} }
It is not difficult to check using \condd\ that
 \eqn\inwe{ F (z_0) = -2S\log C(z_0) + \sum_{k=0}^{\left[{n+1\over
 2}\right]} {C^{2k}\over (k!)^2} W^{(2k)}(z_0) }
satisfies \codo.

Combining the results into \piec, we get our final answer,
$$ I(z_0, \Lambda_0 ) = -W(\Lambda_0 ) + 2 S \log {\Lambda_0\over
C} + \sum_{k=0}^{\left[{n+1\over 2}\right]} {C^{2k}\over (k!)^2}
W^{(2k)}(z_0) + \CO \left( {1\over \Lambda_0}\right). $$

\bigskip

\listrefs

\end